\documentclass[lettersize,journal]{IEEEtran}

\usepackage{amsmath,amsfonts,amssymb}
\usepackage{algorithm}
\usepackage{algorithmic}
\usepackage{array}
\usepackage[caption=false,font=normalsize,labelfont=sf,textfont=sf]{subfig}
\usepackage{textcomp}
\usepackage{stfloats}
\usepackage{url}
\usepackage{verbatim}
\usepackage{graphicx}
\usepackage{cite}
\usepackage{multirow,booktabs}
\usepackage[table,xcdraw]{xcolor}
\usepackage{comment}
\usepackage{pifont}
\usepackage{tablefootnote}
\usepackage{tabularx}
\usepackage{tablefootnote}

\definecolor{baselineblue}{HTML}{E3ECFF}
\definecolor{pairedgray}{HTML}{E5E5E5}

\usepackage[table]{xcolor}
\usepackage{tabularx}
\usepackage{array}
\usepackage{booktabs}
\usepackage{multirow}

\newcolumntype{Y}{>{\centering\arraybackslash}X}
\definecolor{proposedgray}{RGB}{232,232,232}

\providecolor{proposedgray}{gray}{0.9}
\providecolor{baselinegray}{gray}{0.9}

\providecolor{tablegroup}{RGB}{235,239,243}
\providecolor{panelreference}{RGB}{226,230,235}
\providecolor{panelstatic}{RGB}{246,239,219}
\providecolor{panelmemory}{RGB}{220,235,247}
\providecolor{proposedgray}{gray}{0.94}
\providecolor{baselinegray}{gray}{0.91}
\providecolor{baselineblue}{RGB}{226,238,251}

\begin{document}

\title{
Beyond Acoustic Prefixes: Persistent Access to Serialized Acoustic Memory for LLM-Based Multi-Talker Speech Recognition
}

\author{Hao~Shi,~\IEEEmembership{Member,~IEEE,}~Yuan~Gao,~\IEEEmembership{Member,~IEEE,}~Xugang~Lu,~\IEEEmembership{Senior~Member,~IEEE}\\~Tatsuya~Kawahara,~\IEEEmembership{Fellow,~IEEE}
\thanks{Hao~Shi, Yuan~Gao and Tatsuya~Kawahara are with the Graduate School of Informatics, Kyoto University, Kyoto, Japan (email: hshi@ieee.org, yuang@ieee.org, kawahara@i.kyoto-u.ac.jp).}
\thanks{Xugang~Lu is with the National Institute of Information and Communications Technology, Japan (email: xugang.lu@nict.go.jp).}
\thanks{This work has been submitted to the IEEE for possible publication. Copyright may be transferred without notice, after which this version may no longer be accessible.}
}

\markboth{Journal of \LaTeX\ Class Files,~Vol.~14, No.~8, August~2021}%
{Shell \MakeLowercase{\textit{et al.}}: A Sample Article Using IEEEtran.cls for IEEE Journals}

\maketitle

\begin{abstract}
    Large language models (LLMs) provide strong linguistic priors for serialized output training (SOT), yet LLM-based multi-talker ASR degrades substantially as the number of overlapping talkers increases. 
    Conventional systems expose acoustic evidence primarily through an initial projected mixture prefix, requiring the decoder to preserve and recover talker-relevant information indirectly throughout autoregressive generation. 
    We first examine whether this limitation can be resolved by enriching the static prefix using discrete connectionist temporal classification (CTC) tokens, hybrid token--acoustic prompts, and continuous talker-specific representations. 
    The results suggest that acoustic content alone does not fully address the conditioning bottleneck. 
    We therefore extend onset-based serialization from the output target to the acoustic-conditioning pathway and introduce \emph{persistent decoder-side access to SOT-aligned serialized acoustic memory}. 
    Talker-specific representations are organized in utterance-onset order and retained as external acoustic memory, while the conventional mixture prefix provides a complementary global-conditioning path. 
    LLM layers query this memory throughout generation through gated residual cross-attention. 
    We further introduce a second adaptation stage that jointly applies low-rank updates to the acoustic-retrieval pathway and selected LLM self-attention projections. 
    Experiments on LibriMix show consistent improvements over static-prefix prompting.
    These results indicate that effective LLM-based multi-talker ASR depends not only on providing richer acoustic representations, but also on maintaining persistent access to acoustic evidence structured according to the serialized output. 
\end{abstract}


\section{Introduction}
    \IEEEPARstart{M}{ulti-talker} automatic speech recognition (ASR) aims to transcribe speech when multiple talkers speak simultaneously \cite{9054328,cornell24_chime,Cornell2025RecentTrends,zhang2020improving,raj2023surt2}.
    Compared with single-talker ASR \cite{10542371,6732927,6423821,fan2025joint},
    it must recover linguistic content from overlapping voices while resolving which acoustic evidence belongs to each talker \cite{8682822,8461893,9054328}.
    Multi-talker recognition can be approached with an explicit speech-separation frontend \cite{8369155,fan2025crossmodal,6739096}, or by jointly modeling separation, diarization, and recognition \cite{raj2021integration,kanda2021speakerattributed,kanda21_interspeech}. 
    Among joint approaches, permutation-invariant training (PIT)
    \cite{7979557,qian2018singlechannel}
    and serialized output training (SOT) \cite{kanda20b_interspeech,10446059,hirano25_interspeech,kanda22_interspeech,kang2024crossspeaker,kang2025disentangling}
    are two representative paradigms.
    PIT produces multiple output streams and resolves label-permutation ambiguity by minimizing over possible assignments, but requires predefined branches whose assignment and decoding costs grow with the number of talkers.
    SOT instead orders talker transcriptions according to a serialization rule and concatenates them with speaker-change tokens, avoiding multiple autoregressive output streams and representing different output segments within a single sequence~\cite{kanda20b_interspeech}.
    However, recognition, talker assignment, and output ordering must then be solved jointly, making SOT increasingly difficult as overlap and mixture complexity increase \cite{kanda20b_interspeech,LiangYLGZ0023,sakuma2025speakerdistinguishable}.

    Recent work has explored using large language models (LLMs) as ASR decoders \cite{li2023prompting,li24b_interspeech}.
    In multi-talker settings, LLMs have been incorporated through acoustic-prefix conditioning, task instructions, contextual modeling, and diarization-conditioned grounding \cite{shi2025serialized,shi2024advancing,meng2024large,he2025cmt,polok2026grounding,yu2024connecting,meng24c_interspeech}.
    A typical LLM-based ASR system consists of a speech encoder, a temporal-reduction module, a modality projector, and a pretrained LLM decoder.
    The encoded speech sequence is projected into the LLM embedding space and prepended to the text tokens as an acoustic prefix for autoregressive generation \cite{shi2024advancing,meng2024large,yu2024connecting}.
    This formulation is attractive because the LLM supplies a strong linguistic prior that can help resolve locally ambiguous acoustic observations.

    Nevertheless, the main limitation of LLM-based SOT is not that it is uniformly ineffective, but that it scales poorly as the number of talkers increases.
    As observed in our previous work~\cite{shi2025serialized}, LLM-based SOT is competitive on two-talker mixtures but degrades substantially on three-talker mixtures, and increasing the language-decoder capacity does not consistently remove this degradation.
    Although three-or-more-way overlap is less frequent than two-talker overlap, it occurs in real multi-party meetings and exposes a difficult failure case for systems designed around two-speaker concurrency assumptions \cite{laskowski12_interspeech,he2025scaling}. 
    When several speech streams admit linguistically plausible continuations, linguistic compatibility alone cannot determine which talker produced a particular token or token span.
    The decoder must repeatedly associate its serialized output with fine-grained evidence from the corresponding acoustic stream.

    We attribute this talker-count scaling problem to two mismatches between conventional acoustic conditioning and serialized generation.
    The first is a \emph{structural mismatch}.
    SOT organizes the textual target into onset-ordered talker segments, whereas the conventional acoustic prefix remains a mixture-level representation that does not explicitly expose the corresponding talker structure.
    The decoder must therefore infer the relationship between a talker-serialized output and structurally entangled acoustic conditioning.
    The second is an \emph{access-path mismatch}.
    Although talker-specific acoustic evidence may be needed throughout generation, both acoustic retrieval and textual-context modeling are mediated by the same self-attention pathway over the prefix and the growing text history.
    This does not imply that prefix representations become mathematically inaccessible after prefill: their key--value states remain available to later self-attention operations.
    Rather, the mixture prefix provides neither an explicitly talker-aligned organization nor a dedicated pathway for retrieving acoustic evidence during serialized generation.
    Both mismatches become more consequential as the number of possible token-to-talker assignments increases.

    Before changing the conditioning interface, we first ask whether the limitation of prefix-conditioned SOT can be explained simply by insufficient acoustic content.
    Building on our previous serialized output prompting framework~\cite{shi2025serialized}, we revisit three forms of guidance derived from serialized connectionist temporal classification (CTC)
    \cite{shi_slt2024,shi2025serialized}:
    (1) discrete CTC transcription tokens;
    (2) hybrid prompts that combine CTC tokens with projected acoustic representations; and
    (3) continuous prompts constructed from talker-specific CTC hidden representations.
    These variants vary the form and richness of the information supplied to the decoder while retaining static-prefix delivery.
    Acoustic-enriched prompts substantially outperform discrete CTC hypotheses, but their gains over conventional SOT remain modest under three-talker conditions.
    This result suggests that richer acoustic content is helpful but may remain insufficient when it is delivered only through the initial prefix.

    Our central hypothesis is therefore that the limitation of prefix-conditioned SOT is determined not only by how much acoustic information is provided, but also by how that information is structured and accessed.
    We therefore hypothesize that talker-specific representations organized
    according to the SOT order provide more effective acoustic conditioning
    when they remain directly accessible throughout autoregressive decoding. 
    We extend onset-based serialization from the textual output to the acoustic-conditioning pathway. 
    To instantiate this idea, we retain the continuous talker-specific representations in utterance-onset order and expose them as \emph{serialized acoustic memory}.
    The organization of this memory follows the utterance order of the SOT target, establishing order-level correspondence between the acoustic streams and the serialized output.
    Because the memory remains outside the autoregressive token sequence, decoder layers can query it throughout generation.
    We refer to this combination of SOT-aligned serialized memory and a dedicated decoder-side access pathway throughout generation as \emph{SOT-aligned persistent acoustic access}.

    The conventional projected mixture prefix is retained, yielding a dual-path conditioning interface: the prefix supplies global mixture context through self-attention, while decoder layers in the LLM retrieve talker-specific evidence from the serialized acoustic memory through cross-attention throughout generation.
    A learnable gate controls the magnitude of the retrieved residual update.
    Although external modality memory, cross-attention retrieval, and gated or zero-initialized adapters have substantial precedents in multimodal LLMs \cite{alayrac2022flamingo,zhang2024llamaadapter,li2023blip2},
    we use these mechanisms to realize an SOT-specific conditioning principle: preserving onset-ordered acoustic evidence as an external memory that remains accessible during serialized generation.
    The contribution therefore lies in this serialization-aligned persistent-memory formulation and its controlled validation, rather than in cross-attention or gating alone.
    Prefix conditioning and external-memory retrieval impose different acoustic-access biases rather than different theoretical limits on decoder expressiveness.

    To validate the proposed formulation and determine which of its properties are essential, we organize the evaluation around three groups of controlled comparisons.
    First, a fully crossed comparison varies the memory source---mixture-level or onset-ordered talker-specific representations---and the residual integration mechanism---ungated addition or learnable gating---under the same initialization and optimization budget.
    Second, we compare a dual-static-prefix system with a persistent-memory system using the same serialized acoustic source, projector, and temporal resolution.
    A separate prefill-only control uses the same cross-attention module as the persistent system but disables memory access during autoregressive generation, thereby isolating the contribution of continued decoder-time retrieval from additional module capacity.
    Third, we remove the conventional mixture prefix while retaining access to the serialized talker memory, thereby isolating the contribution of the complementary dual-path design.
    Together, these comparisons examine the respective contributions of talker-aligned memory structure, persistent access, residual gating, and dual-path conditioning.

    For the complete system, we further propose a two-stage adaptation strategy.
    The first stage learns the gated acoustic-memory pathway while preserving the pretrained LLM weights.
    The resulting gated model initializes a second stage in which low-rank adaptation (LoRA) is applied jointly to the trained cross-attention projections and selected LLM self-attention projections~\cite{hu2021lora}.
    This second stage is treated as a downstream optimization of an already established memory interface and is evaluated separately from the controlled architectural comparisons, so that its effect is not conflated with memory structure, access interface, or additional training budget.

\section{Preliminaries}
\label{sec:preliminaries}

    \subsection{LLM-Based Serialized Multi-Talker ASR}
    \label{sec:llm_asr_prelim}

        Given an input waveform $\mathbf{y}$, a speech encoder, temporal-reduction module, and modality projector produce a mixture-level acoustic prefix
        \begin{equation}
        \setlength{\abovedisplayskip}{2pt}
        \setlength{\belowdisplayskip}{2pt}
        \mathbf{H}_e=\mathrm{Enc}(\mathbf{y}),
        \qquad
        \mathbf{H}_p
        =
        \mathrm{Projector}_{\mathrm{mix}}
        \left(\mathrm{Down}(\mathbf{H}_e)\right),
        \label{eq:prefix_encoding}
        \end{equation}
        where $\mathbf{H}_e\in\mathbb{R}^{T_e\times D_e}$ and
        $\mathbf{H}_p\in\mathbb{R}^{T_p\times D_{\mathrm{model}}}$.
        Here, $T_e$ and $D_e$ denote the encoder-sequence length and encoder hidden
        dimension, respectively; $T_p$ is the acoustic-prefix length after temporal
        reduction; and $D_{\mathrm{model}}$ is the LLM hidden dimension.
        $\mathrm{Down}(\cdot)$ denotes temporal reduction, and
        $\mathrm{Projector}_{\mathrm{mix}}(\cdot)$ is a learned projection into the
        LLM hidden space.
        The conventional LLM input is
        \begin{equation}
        \setlength{\abovedisplayskip}{2pt}
        \setlength{\belowdisplayskip}{2pt}
        \mathbf{X}
        =
        \operatorname{Concat}
        \left(
        \mathbf{E}_{\mathrm{inst}},
        \mathbf{H}_p,
        \mathbf{E}_t
        \right),
        \label{input_sequence}
        \end{equation}
        where $\mathbf{E}_{\mathrm{inst}}$ is an optional instruction embedding and
        $\mathbf{E}_t$ contains the right-shifted text embeddings.
        Throughout the paper, $\operatorname{Concat}(\cdot)$ denotes concatenation
        along the sequence (time) axis; all concatenated representations therefore
        share the feature dimension $D_{\mathrm{model}}$.
        For example, for the instruction-tuned LLM, we use the instruction ``TRANSCRIBE THE PROVIDED AUDIO INTO ACCURATE TEXT.'' Its embedding sequence is denoted by $\mathbf{E}_{\mathrm{inst}}$ and prepended to the acoustic input. For the base LLM variants, $\mathbf{E}_{\mathrm{inst}}$ is omitted. 
        We call $\mathbf{H}_p$ a \emph{static prefix} because it is placed in the autoregressive input rather than retained behind a separate decoder-side memory interface; its cached key--value states nevertheless remain accessible to later self-attention operations.
        The encoder and LLM can be initialized from pretrained models~\cite{9814838,10096630,radford2023robust}, and the LLM can be adapted using parameter-efficient methods such as LoRA~\cite{hu2021lora}.

        SOT converts multi-talker recognition into single-sequence generation~\cite{kanda20b_interspeech}.
        For $K$ talkers ordered by utterance onset, with transcription $\mathbf{t}_k$ for the $k$-th talker in that order, the target is
        \begin{equation}
        \setlength{\abovedisplayskip}{2pt}
        \setlength{\belowdisplayskip}{2pt}
        \mathbf{T}_{\mathrm{sot}}
        =
        \operatorname{Concat}
        \left(
        \mathbf{t}_1,
        \langle\mathrm{sc}\rangle,
        \mathbf{t}_2,
        \ldots,
        \langle\mathrm{sc}\rangle,
        \mathbf{t}_K
        \right),
        \label{eq:sot_target}
        \end{equation}
        where $\langle\mathrm{sc}\rangle$ is a single speaker-change token type reused
        between consecutive talker segments and therefore occurs $K-1$ times in
        a $K$-talker target.
        The token marks segment boundaries rather than speaker identities;
        the permutation ambiguity is resolved by ordering the segments according
        to utterance onset.
        The LLM is trained autoregressively using
        \begin{equation}
        \setlength{\abovedisplayskip}{2pt}
        \setlength{\belowdisplayskip}{2pt}
        \mathcal{L}_{\mathrm{SOT}}
        =
        \mathrm{CE}
        \left(
        \mathbf{T}_{\mathrm{sot}},
        \mathbf{Z}_{\mathrm{sot}}
        \right),
        \label{eq:sot_ce}
        \end{equation}
        where $\mathbf{Z}_{\mathrm{sot}}$ denotes the LLM output logits and
        $\mathrm{CE}(\cdot,\cdot)$ denotes token-level cross-entropy between the
        serialized target and the corresponding autoregressive output distribution.

    \subsection{Serialized CTC and Previous Static Prompting}
    \label{sec:gencsep}

        Our previous EncSep framework~\cite{shi_slt2024} introduced an encoder-side separator that produces $K$ talker-specific streams from the shared acoustic representation:
        \begin{equation}
        \setlength{\abovedisplayskip}{2pt}
        \setlength{\belowdisplayskip}{2pt}
        \left\{
        \mathbf{H}_{\mathrm{sep}}^k
        \right\}_{k=1}^{K}
        =
        \mathrm{Separator}
        \left(
        \mathbf{H}_e
        \right).
        \label{eq:separator_output}
        \end{equation}
        The separator consisted of a long short-term memory (LSTM) network followed by layer normalization and talker-dependent projections; its exact configuration is reported in Section~\ref{sec:model_configs}.
        Each stream is mapped to CTC logits
        $\mathbf{Z}_{\mathrm{ctc}}^k=\mathrm{CTCHead}(\mathbf{H}_{\mathrm{sep}}^k)$ and supervised by the transcription with the same onset-order index:
        \begin{equation}
        \setlength{\abovedisplayskip}{2pt}
        \setlength{\belowdisplayskip}{2pt}
        \mathcal{L}_{\mathrm{SerCTC}}
        =
        \sum_{k=1}^{K}
        \mathrm{CTC}
        \left(
        \mathbf{Z}_{\mathrm{ctc}}^k,
        \mathbf{t}_k
        \right).
        \label{serialized-ctc-loss}
        \end{equation}
        Here, $\mathrm{CTC}(\cdot,\cdot)$ denotes the connectionist temporal
        classification loss between a talker-specific logit sequence and its
        onset-ordered transcription.
        This supervision aligns the separator branch index with the SOT output order.

        Our SOP framework~\cite{shi2025serialized} converted the greedy serialized-CTC hypotheses into token embeddings and concatenated them with the downsampled mixture prefix before autoregressive decoding.
        Specifically, if
        $\bar{\mathbf{C}}_{\mathrm{ser}}$
        denotes the onset-ordered concatenation of the talker-level CTC hypotheses, SOP used
        \begin{equation}
        \setlength{\abovedisplayskip}{2pt}
        \setlength{\belowdisplayskip}{2pt}
        \mathbf{X}_{\mathrm{sop}}
        =
        \operatorname{Concat}
        \left(
        \mathbf{E}_{\mathrm{inst}},
        \mathrm{Embedding}(\bar{\mathbf{C}}_{\mathrm{ser}}),
        \mathbf{H}_p,
        \mathbf{E}_t
        \right).
        \label{eq:sop_input}
        \end{equation}
        Thus, EncSep learned onset-ordered talker representations, whereas SOP exposed their decoded hypotheses to the LLM only through the initial input sequence.
        The next section retains these essential components but compares several CTC-derived static representations more directly.

    \section{Characterizing Static Prefix Conditioning with CTC-Derived Cues}
    \label{sec:revisit_prefix}

        Before introducing decoder-side memory, we examine how discrete and continuous serialized-CTC cues behave when delivered only through the initial LLM input.
        These variants characterize different static information sources but are not, by themselves, single-factor interface controls because they differ in whether the conventional mixture prefix is retained.
        We therefore also construct a matched dual-prefix baseline for comparison with persistent access.

    \subsection{Shared Talker-Specific Representations}
    \label{sec:encoder_side_extraction}

        Unlike the SOP system, which attached its separator to an intermediate downsampling representation, the present systems apply the separator in Eq.~(\ref{eq:separator_output}) directly to $\mathbf{H}_e$ while retaining the conventional mixture-prefix pathway in Eq.~(\ref{eq:prefix_encoding}).
        Greedy CTC decoding is performed independently for each onset-ordered branch, and the resulting hypotheses are concatenated as
        \begin{equation}
        \setlength{\abovedisplayskip}{2pt}
        \setlength{\belowdisplayskip}{2pt}
        \bar{\mathbf{C}}_{\mathrm{ser}}
        =
        \operatorname{Concat}
        \left(
        \mathrm{GreedyCTC}(\mathbf{Z}_{\mathrm{ctc}}^1),
        \ldots,
        \mathrm{GreedyCTC}(\mathbf{Z}_{\mathrm{ctc}}^K)
        \right).
        \label{eq:serialized_ctc_tokens}
        \end{equation}
        Repeated symbols and CTC blanks are removed by greedy decoding. 
        No speaker-change token is inserted between the
        talker-level hypotheses before concatenation. 
        The corresponding continuous streams are concatenated in the same order:
        \begin{equation}
        \setlength{\abovedisplayskip}{2pt}
        \setlength{\belowdisplayskip}{2pt}
        \mathbf{H}_{\mathrm{ser}}
        =
        \operatorname{Concat}
        \left(
        \mathbf{H}_{\mathrm{sep}}^1,
        \ldots,
        \mathbf{H}_{\mathrm{sep}}^K
        \right).
        \label{eq:serialized_acoustic_features}
        \end{equation}
        Unless otherwise stated, the static and persistent systems use separator and serialized-CTC components initialized from the same CTC-stage checkpoint.

    \subsection{CTC-Derived Static Prefixes}
    \label{sec:prompt_variants}

        We consider three static variants derived from Eqs.~(\ref{eq:serialized_ctc_tokens}) and~(\ref{eq:serialized_acoustic_features}):
        \begin{equation}
        \setlength{\abovedisplayskip}{2pt}
        \setlength{\belowdisplayskip}{2pt}
        \begin{aligned}
        \mathbf{E}_{\mathrm{tok}}
        &=
        \mathrm{Embedding}(\bar{\mathbf{C}}_{\mathrm{ser}}),\\
        \mathbf{E}_{\mathrm{hyb}}
        &=
        \operatorname{Concat}(\mathbf{E}_{\mathrm{tok}},\mathbf{H}_p),\\
        \mathbf{E}_{\mathrm{aco}}
        &=
        \mathrm{Projector}_{\mathrm{sep}}(\mathbf{H}_{\mathrm{ser}}).
        \end{aligned}
        \label{eq:static_prompt_representations}
        \end{equation}
        Here, $\mathrm{Embedding}(\cdot)$ denotes the LLM token-embedding lookup,
        and $\mathrm{Projector}_{\mathrm{sep}}(\cdot)$ is a learned mapping from the
        separator feature dimension to $D_{\mathrm{model}}$.
        TokenPrompt uses $\mathbf{E}_{\mathrm{tok}}$ and therefore supplies talker-ordered lexical hypotheses without frame-level acoustic states.
        HybridPrompt uses $\mathbf{E}_{\mathrm{hyb}}$ and retains the conventional mixture prefix in addition to those hypotheses.
        AcousticPrompt uses $\mathbf{E}_{\mathrm{aco}}$, preserving continuous talker-specific information before the CTC output projection, but replaces rather than augments $\mathbf{H}_p$.
        Their inputs can be written compactly as
        \begin{equation}
        \setlength{\abovedisplayskip}{2pt}
        \setlength{\belowdisplayskip}{2pt}
        \mathbf{X}_{v}
        =
        \operatorname{Concat}
        \left(
        \mathbf{E}_{\mathrm{inst}},
        \mathbf{E}_{v},
        \mathbf{E}_t
        \right),
        \quad
        \mathbf{E}_{v}
        \in
        \left\{
        \mathbf{E}_{\mathrm{tok}},
        \mathbf{E}_{\mathrm{hyb}},
        \mathbf{E}_{\mathrm{aco}}
        \right\}.
        \label{eq:prompt_variants}
        \end{equation}
        All three variants are optimized using the SOT objective in Eq.~(\ref{eq:sot_ce}).

    \subsection{Matched Dual-Prefix Baseline}
    \label{sec:matched_dual_prefix}

        AcousticPrompt is not an interface-matched control for the proposed system because it removes $\mathbf{H}_p$ and uses a separate static projection pathway.
        We therefore construct a dual-prefix baseline containing both the conventional mixture prefix and the same serialized representation used by its persistent counterpart:
        \begin{equation}
        \setlength{\abovedisplayskip}{2pt}
        \setlength{\belowdisplayskip}{2pt}
        \mathbf{M}_{\mathrm{match}}
        =
        \mathrm{Projector}_{\mathrm{match}}
        \left(
        \mathrm{Reduce}_{\mathrm{match}}
        (\mathbf{H}_{\mathrm{ser}})
        \right).
        \label{eq:matched_serialized_representation}
        \end{equation}
        The matched static and persistent systems share this acoustic source, temporal reduction, and projector.
        Here, $\mathrm{Reduce}_{\mathrm{match}}(\cdot)$ denotes the common matched
        temporal-reduction operator, and $\mathrm{Projector}_{\mathrm{match}}(\cdot)$
        denotes the common learned projection into the LLM hidden space.
        The static system concatenates it into the LLM input:
        \begin{equation}
        \setlength{\abovedisplayskip}{2pt}
        \setlength{\belowdisplayskip}{2pt}
        \mathbf{X}_{\mathrm{dual}}
        =
        \operatorname{Concat}
        \left(
        \mathbf{E}_{\mathrm{inst}},
        \mathbf{H}_p,
        \mathbf{M}_{\mathrm{match}},
        \mathbf{E}_t
        \right),
        \label{eq:matched_dual_prefix}
        \end{equation}
        whereas the matched persistent system retains $\mathbf{M}_{\mathrm{match}}$ outside the token sequence and supplies it through decoder-side cross-attention.

\begin{figure}[t]
    \centering
    \includegraphics[width=0.9\linewidth]{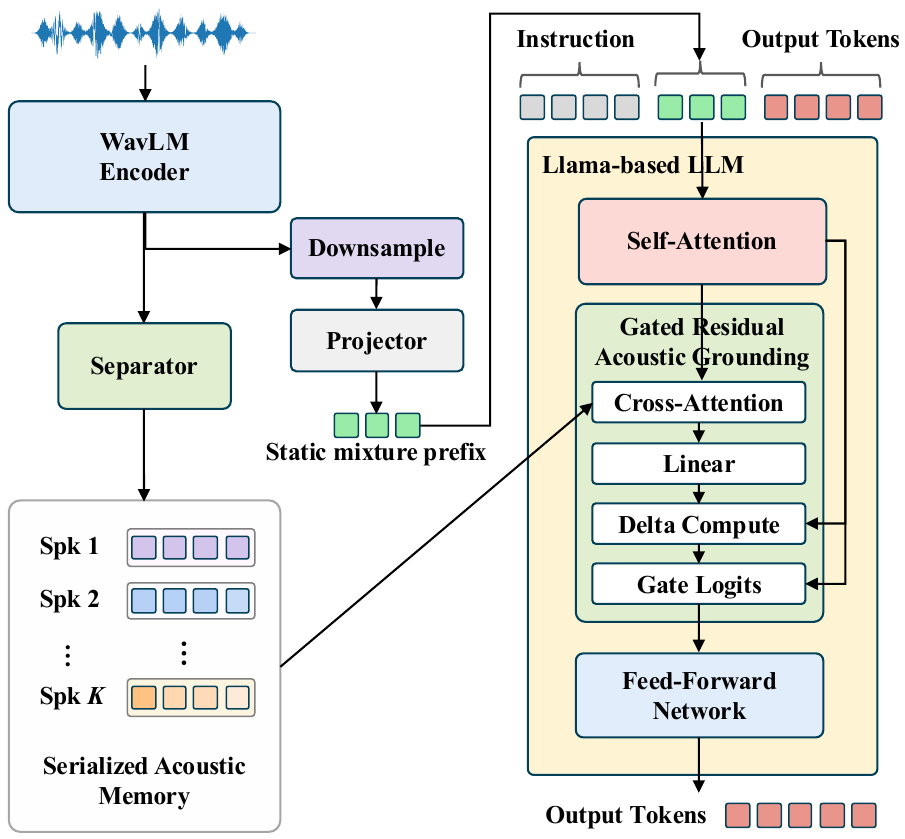}
    \vspace{-1mm}
    {\footnotesize
    Separator/CTC heads $\rightarrow\mathcal{L}_{\mathrm{SerCTC}}$
    \qquad
    LLM output logits $\rightarrow\mathcal{L}_{\mathrm{SOT}}$\par}
    \caption{\textbf{SOT-aligned persistent acoustic access.}
    Continuous talker-specific representations are serialized according to
    utterance onset and retained as external acoustic memory.
    LLM layers are equipped with residual cross-attention to this
    memory during prefill and autoregressive decoding, while the conventional
    mixture prefix provides complementary static conditioning in the complete
    dual-path system.
    }
    \label{fig:network}
\end{figure}

\section{Persistent Access to Serialized Acoustic Memory}
\label{sec:proposed_method}

    The preceding static variants represent all acoustic and CTC-derived evidence within the autoregressive input sequence.
    Their acoustic states remain available through self-attention, but share the same attention pathway as the textual context.
    We introduce persistent access to serialized acoustic memory to study a different access pattern: continuous onset-ordered talker representations are retained outside the autoregressive token stream and supplied to selected decoder layers through a dedicated cross-attention pathway.
    This changes where the acoustic evidence is stored and how decoder states access it during generation. 
    The main system retains the conventional mixture prefix and additionally exposes serialized talker representations as external memory.
    This dual-path construction is evaluated against matched static, prefill-only, and memory-only controls so that the effects of acoustic source, continued access, and mixture-prefix conditioning can be separated.
    We further introduce a two-stage acoustic adaptation strategy tailored to this persistent-memory interface: a decoder-side retrieval pathway is learned first, after which low-rank updates are introduced into the trained cross-attention projections and selected LLM self-attention projections.

    \subsection{Serialized Acoustic Memory}
    \label{sec:serialized_acoustic_memory}
    
        The proposed method uses the continuous sequence $\mathbf{H}_{\mathrm{ser}}$ in Eq.~(\ref{eq:serialized_acoustic_features}) rather than decoded CTC hypotheses.
        A memory projector maps the serialized talker representation to the LLM hidden dimension:
        \begin{equation}
        \setlength{\abovedisplayskip}{2pt}
        \setlength{\belowdisplayskip}{2pt}
        \mathbf{M}_{\mathrm{ser}}
        =
        \mathrm{Projector}_{\mathrm{mem}}
        \left(
        \mathbf{H}_{\mathrm{ser}}
        \right).
        \label{eq:serialized_memory}
        \end{equation}
        Here, $\mathbf{M}_{\mathrm{ser}}\in\mathbb{R}^{T_m\times D_{\mathrm{model}}}$ is the serialized acoustic memory, where $T_m$ is its length.
        The individual streams are arranged according to the same onset rule used for the SOT target, so the order of the memory blocks follows the order of the output segments.

        The conventional decoder input of the main system remains
        $\mathbf{X}_{\mathrm{main}}=\mathbf{X}$, as defined in
        Eq.~(\ref{input_sequence}),
        while $\mathbf{M}_{\mathrm{ser}}$ is provided separately to the adapted decoder layers.
        Thus, $\mathbf{H}_p$ and $\mathbf{M}_{\mathrm{ser}}$ provide two complementary conditioning paths: a mixture-level input prefix and a decoder-side talker-structured memory.

        For the memory-source control, we also construct mixture-level memory from the encoder sequence:
        \begin{equation}
        \setlength{\abovedisplayskip}{2pt}
        \setlength{\belowdisplayskip}{2pt}
        \mathbf{M}_{\mathrm{mix}}
        =
        \mathrm{Projector}_{\mathrm{mem}}^{\mathrm{mix}}
        \left(
        \mathbf{H}_e
        \right).
        \label{eq:mixture_memory}
        \end{equation}
        The mixture and serialized memory projectors use the same output dimension and matched projector architecture.
        This permits the memory source to be varied without changing the decoder-side cross-attention formulation.

    \subsection{Decoder-Side Acoustic Retrieval}
    \label{sec:decoder_side_retrieval}

        We insert a lightweight cross-attention sublayer after the self-attention sublayer of each LLM layer.
        Let $\mathbf{H}^{(\ell)}\in\mathbb{R}^{L_t\times D_{\mathrm{model}}}$ denote the hidden states after self-attention in layer $\ell$, where $L_t$ is the current decoder-sequence length.
        For an acoustic memory
        $\mathbf{M}_s\in\mathbb{R}^{T_s\times D_{\mathrm{model}}}$, with
        $s\in\{\mathrm{mix},\mathrm{ser}\}$ and memory length $T_s$, the query, key,
        and value representations are computed as
        \begin{equation}
        \setlength{\abovedisplayskip}{2pt}
        \setlength{\belowdisplayskip}{2pt}
        \begin{split}
        \mathbf{Q}^{(\ell)}
        =
        \operatorname{LayerNorm}_{\mathrm{in}}^{(\ell)}
        (\mathbf{H}^{(\ell)})\mathbf{W}_{q}^{(\ell)},\\
        [\mathbf{K}^{(\ell)}\;\mathbf{V}^{(\ell)}]
        =
        \mathbf{M}_s
        [\mathbf{W}_{k}^{(\ell)}\;\mathbf{W}_{v}^{(\ell)}].
        \end{split}
        \label{eq:retrieval_qkv}
        \end{equation}
        Here, $\operatorname{LayerNorm}_{\mathrm{in}}^{(\ell)}(\cdot)$ is the input
        LayerNorm of the cross-attention sublayer.  Let $D_a$ denote the total
        cross-attention projection dimension.  The learned matrices $\mathbf{W}_{q}^{(\ell)},\mathbf{W}_{k}^{(\ell)},\mathbf{W}_{v}^{(\ell)}
        \in\mathbb{R}^{D_{\mathrm{model}}\times D_a}$ project decoder or memory states
        to queries, keys, and values, respectively, and
        $\mathbf{W}_{o}^{(\ell)}\in\mathbb{R}^{D_a\times D_{\mathrm{model}}}$ projects
        the retrieved context back to the LLM hidden dimension.

        For multi-head attention, let $d_h$ denote the dimensionality of each attention head.
        The complete retrieved acoustic update is
        \begin{equation}
        \setlength{\abovedisplayskip}{2pt}
        \setlength{\belowdisplayskip}{2pt}
        \mathbf{U}^{(\ell)}
        =
        \mathrm{Softmax}
        \left(
        \frac{
        \mathbf{Q}^{(\ell)}
        \mathbf{K}^{(\ell)\top}
        }{
        \sqrt{d_h}
        }
        +
        \mathbf{S}
        \right)
        \mathbf{V}^{(\ell)}
        \mathbf{W}_{o}^{(\ell)},
        \label{eq:retrieved_acoustic_update}
        \end{equation}
        where $\mathbf{S}\in\mathbb{R}^{L_t\times T_s}$ is a masking bias that excludes padded memory positions and is broadcast across attention heads.
        The ``Delta Compute'' block in Fig.~\ref{fig:network} denotes the computation
        of this retrieved acoustic update $\mathbf{U}^{(\ell)}$.
        
        We consider stacked and gated residual interfaces under the unified formulation
        \begin{equation}
        \setlength{\abovedisplayskip}{2pt}
        \setlength{\belowdisplayskip}{2pt}
        \mathbf{H}_{r}^{(\ell)}
        =
        \mathbf{H}^{(\ell)}
        +
        \eta_{\ell,r}\mathbf{U}^{(\ell)},
        \eta_{\ell,\mathrm{stack}}=1,
        \eta_{\ell,\mathrm{gate}}=\sigma(\gamma_{\ell}),
        \label{eq:retrieval_residual_interfaces}
        \end{equation}
        where $r\in\{\mathrm{stack},\mathrm{gate}\}$, $\gamma_{\ell}$ is a trainable
        scalar gate logit, and $\sigma(\cdot)$ denotes the sigmoid function.
        Thus, the stacked interface directly adds the retrieved update, whereas the gated interface explicitly controls its magnitude.
        The resulting states are then passed through the original feed-forward sublayer of the LLM.

    \subsection{Two-Stage Acoustic Adaptation}
    \label{sec:two_stage_adaptation}

        We introduce a two-stage acoustic adaptation strategy tailored to the persistent-memory interface.
        The two stages refer specifically to decoder-side adaptation after the separator and serialized CTC branches have been initialized.

        \paragraph{Stage 1: acoustic-memory interface}
            In the first stage, the dense parameters of the pretrained LLM are kept fixed while the memory projector and decoder-side cross-attention parameters are optimized.
            For a gated configuration, the gate logits are optimized in the same stage.
            This produces a trained cross-attention pathway without modifying the dense pretrained LLM weights.

        \paragraph{Stage 2: joint low-rank refinement}
            The trained Stage~1 cross-attention model initializes the second stage.
            All dense projection matrices remain fixed, and LoRA modules are introduced into selected LLM self-attention projections and the query, key, value, and output projections of cross-attention.
            Let $\mathcal{W}_{\mathrm{sa}}^{(\ell)}$ denote the selected self-attention matrices and let
            \begin{equation}
            \setlength{\abovedisplayskip}{2pt}
            \setlength{\belowdisplayskip}{2pt}
            \mathcal{W}_{\mathrm{ca}}^{(\ell)}
            =
            \left\{
            \mathbf{W}_{q}^{(\ell)},
            \mathbf{W}_{k}^{(\ell)},
            \mathbf{W}_{v}^{(\ell)},
            \mathbf{W}_{o}^{(\ell)}
            \right\}
            \label{eq:cross_attention_weight_set}
            \end{equation}
            denote the cross-attention matrices.
            For each
            $\mathbf{W}^{(\ell)}\in
            \mathcal{W}_{\mathrm{sa}}^{(\ell)}
            \cup
            \mathcal{W}_{\mathrm{ca}}^{(\ell)}$,
            we introduce
            \begin{equation}
            \setlength{\abovedisplayskip}{2pt}
            \setlength{\belowdisplayskip}{2pt}
            \widetilde{\mathbf{W}}^{(\ell)}
            =
            \mathbf{W}^{(\ell)}
            +
            \frac{\alpha_{\mathrm{LoRA}}}{r}
            \mathbf{B}^{(\ell)}
            \mathbf{A}^{(\ell)}.
            \label{eq:joint_lora_refinement}
            \end{equation}
            For a projection
            $\mathbf{W}^{(\ell)}\in\mathbb{R}^{d_{\mathrm{out}}\times d_{\mathrm{in}}}$,
            the trainable factors satisfy
            $\mathbf{A}^{(\ell)}\in\mathbb{R}^{r\times d_{\mathrm{in}}}$ and
            $\mathbf{B}^{(\ell)}\in\mathbb{R}^{d_{\mathrm{out}}\times r}$.
            The rank $r$ controls the update capacity, and $\alpha_{\mathrm{LoRA}}$ controls its scale.
            This stage introduces trainable low-rank perturbations on both sides of the decoder-side retrieval interface without introducing another source of acoustic information.
            Its incremental contribution is evaluated against the corresponding pre-LoRA Stage~1 model and is not used to infer the effects of memory structure or access pattern.

        After training, the low-rank updates are merged into their corresponding projection matrices:
        \begin{equation}
        \setlength{\abovedisplayskip}{2pt}
        \setlength{\belowdisplayskip}{2pt}
        \mathbf{W}^{(\ell)}
        \leftarrow
        \widetilde{\mathbf{W}}^{(\ell)}.
        \label{eq:lora_merge}
        \end{equation}
        Consequently, LoRA introduces no additional inference-time branches.
        The cross-attention sublayers remain part of the decoder because they implement access to the external acoustic memory.

    \subsection{Training Objectives}
    \label{sec:proposed_training_objectives}

        Before the two-stage acoustic adaptation, the encoder and SOT decoder were first trained using $\mathcal{L}_{\mathrm{SOT}}$.
        The SOT pathway was then frozen while the separator and talker-specific CTC branches were initialized using $\mathcal{L}_{\mathrm{SerCTC}}$ in Eq.~(\ref{serialized-ctc-loss}).
        After this initialization, the trained encoder, separator, and CTC-related components were frozen and used to construct the serialized acoustic memory.
        Both Stage~1 and Stage~2 are optimized exclusively with the SOT objective: 
        \begin{equation}
        \setlength{\abovedisplayskip}{2pt}
        \setlength{\belowdisplayskip}{2pt}
        \mathcal{L}_{\mathrm{Stage\text{-}1}}
        =
        \mathcal{L}_{\mathrm{Stage\text{-}2}}
        =
        \mathcal{L}_{\mathrm{SOT}}.
        \label{eq:decoder_adaptation_loss}
        \end{equation}
        Under this common objective, Stage~1 updates the memory projector and cross-attention parameters while keeping the dense pretrained LLM fixed.
        Stage~2 starts from the trained Stage~1 model, keeps its dense parameters fixed, and optimizes only the newly introduced low-rank updates in the selected self-attention and cross-attention projections.

\section{Experimental Setup}
\label{sec:experimental_setup}

    \subsection{Datasets and Evaluation Protocol}
    \label{sec:datasets}

        We conducted experiments on LibriMix~\cite{cosentino2020librimix}, a widely used overlapped-speech benchmark constructed from LibriSpeech~\cite{7178964}.
        We evaluated two- and three-talker recognition using Libri2Mix and Libri3Mix, respectively. 
        Source onset times were controlled using offset files.
        For Libri2Mix, we adopted the offsets from the ESPnet SOT recipe\footnote{\url{https://github.com/espnet/espnet/tree/master/egs2/librimix/sot_asr1}}; for Libri3Mix, we created additional configurations to obtain diverse three-talker onset patterns.
        The Libri3Mix offset configurations are publicly available in our released source code\footnote{\url{https://github.com/hshi-speech/LibriMix-repo/tree/main/Libri3Mix_offset}}. 
        These offsets introduced delays between source utterances rather than prescribing a fixed overlap ratio. 
        Using speaker-activity labels at a 20-ms frame resolution, we computed overlap as the number of frames with at least two active speakers divided by the number of frames with at least one active speaker. 
        Frame counts were aggregated across each test set before division.
        The resulting overlap ratios were approximately 40.2\% and 55.2\% for the offset-controlled Libri2Mix and Libri3Mix test sets, respectively. 
        For Libri3Mix, three speakers were simultaneously active in approximately 17.6\% of speech-active frames. 
        Models in Exp.~1 to Exp.~47 were trained separately on Libri2Mix Clean, Libri2Mix Noisy,
        Libri3Mix Clean, and Libri3Mix Noisy, respectively, and each was evaluated
        on the development and test sets matching its training condition.

        For cross-domain evaluation, we additionally used LibriCSS \cite{9053426}, a far-field multi-talker corpus recorded in a real meeting room. 
        Our models assumed a known talker count and used separate encoder-side configurations for two- and three-talker recognition. 
        We therefore derived two fixed-count evaluation sets from the official LibriCSS release; the segmentation script and complete clip and reference lists are publicly available\footnote{\url{https://github.com/hshi-speech/grounded-multitalker-asr/tree/main/libricss}}. 
        For each session, the utterances in the official meeting transcription were sorted by start time and independently partitioned into consecutive, non-overlapping blocks of two or three utterances. 
        Audio was extracted from channel~0 of the seven-channel array over this interval without additional padding. 
        This procedure produced 1,754 two-talker clips containing 62,579 reference words and 508 three-talker clips containing 23,719 reference words. 
        The two- and three-talker sets contained 5.7 and 2.1 hours of
        audio, respectively, with mean (median/90th-percentile) clip
        durations of 11.8 (11.5/18.0) s and 14.7 (15.2/19.2) s. 
        Both sets covered the six official LibriCSS conditions: non-overlapping speech with short or long inter-utterance silence (0S and 0L) and nominal overlap ratios of 10\%, 20\%, 30\%, and 40\%.
        Each reference was formed by lowercasing the official utterance transcriptions and concatenating them with the speaker-change token in utterance-onset order. 
        We applied the corresponding Libri2Mix- or Libri3Mix-trained model without adaptation to LibriCSS and computed WER directly against this onset-serialized SOT reference, without permutation search or oracle speaker reassignment.

        We reported word error rate (WER), where lower values indicated better recognition performance. 
        For each mixture, the reference transcriptions were concatenated according to utterance onset order using the same serialized target construction as in SOT training. 
        The generated serialized transcription was scored directly against this onset-ordered reference without permutation search or oracle speaker reassignment. 
        Consequently, the reported WER reflected both recognition accuracy and consistency with the prescribed serialization order. 
        Speaker-change tokens were used only as serialization delimiters and were excluded from word-level scoring.

    \subsection{Potential Text Contamination}
        LibriMix and LibriCSS were derived from LibriSpeech utterances whose transcripts originated from public-domain audiobooks.
        We therefore could not completely rule out overlap between their textual content and the pretraining corpora of LLaMA-family models, and the absolute WERs of LLM-based systems on these benchmarks should be interpreted with this caveat.
        Our principal comparisons nevertheless used the same pretrained LLM backbone and the same evaluation data, and differed only in the acoustic source, conditioning interface, or adaptation strategy under study. 
        Moreover, the most challenging noisy and three-talker conditions required acoustic stream assignment and serialized ordering in addition to predicting linguistically plausible text.

    \subsection{Model Configurations}
    \label{sec:model_configs}

        \paragraph{Speech encoder and LLM backbones}
            All systems were implemented using Hugging Face Transformers. 
            The input consisted of 16-kHz waveforms normalized using the Wav2Vec2 feature extractor. 
            We used the 24-layer WavLM-Large model as the speech encoder. 
            Its output representation had a dimension of 1,024 and a frame rate of 50~Hz, corresponding to one frame every 20~ms. 
            For the conventional mixture-prefix pathway, three stride-2 convolutional adapter layers reduced the temporal resolution by a factor of eight to approximately 6.25~Hz. 
            The resulting sequence was mapped to the LLM hidden dimension by the modality projector. 
            We evaluated LLaMA decoders at three scales: LLaMA-3.2-1B, LLaMA-3.2-3B, and LLaMA-3.1-8B, together with their instruction-tuned counterparts. 
            For the instruction-tuned models, we prepended the fixed instruction ``TRANSCRIBE THE PROVIDED AUDIO INTO ACCURATE TEXT,'' whose embedding sequence was denoted by $\mathbf{E}_{\mathrm{inst}}$.

        \paragraph{Encoder-side separator and serialized CTC}
            Following EncSep~\cite{shi_slt2024}, the separator first applied a linear projection from 1,024 to 796 dimensions, followed by a rectified linear unit and layer normalization. 
            The projected sequence was then processed by a two-layer LSTM with a hidden size of 796 and dropout of 0.2. 
            A subsequent layer normalization and $K$ talker-specific branches mapped the LSTM output back to 1,024 dimensions. 
            Each branch consisted of two linear transformations, $796\rightarrow796\rightarrow1{,}024$, with a rectified linear unit and layer normalization. 
            Here, $K$ was fixed to two for Libri2Mix and three for Libri3Mix. 
            Each separated stream was connected to an independent CTC head. 
            The CTC head was a linear transformation from 1,024 dimensions to 128,265 output classes, including the additional CTC blank symbol. 
            The remaining classes followed the non-blank vocabulary of the corresponding LLM tokenizer.

        \paragraph{Serialized acoustic memory and gated decoder-side access}
            The talker-specific representations were concatenated in utterance-onset order without temporal downsampling. 
            Consequently, the serialized acoustic memory retained the native 50-Hz WavLM resolution and had a length of $T_m=K T_e$, where $T_e$ denotes the number of WavLM frames. 
            Each decoder layer contained one independent single-head cross-attention adapter placed after the self-attention residual and before the feed-forward network. 
            Each adapter used an attention bottleneck dimension of $D_a=512$. 
            The decoder hidden states were mapped to queries, while the 1,024-dimensional memory states were mapped to keys and values.
            The retrieved context was then projected back to the LLM hidden dimension. 
            The adapter additionally contained input and output layer-normalization modules and used an attention dropout of 0.0. 
            Padded memory positions were excluded using an attention mask. 
            For gated access, the residual contribution was scaled by $g_{\ell}=\sigma(\gamma_{\ell})$, with one scalar gate for each decoder layer.


    \subsection{Initialization and Training Pipeline}
    \label{sec:training_details}


        \paragraph{SOT backbone training}
            We first trained an LLM-based SOT model using only $\mathcal{L}_{\mathrm{SOT}}$. 
            The WavLM encoder, temporal-reduction module, and modality projector were trainable, whereas the dense LLM parameters remained frozen. 
            LoRA was applied to the query, key, value, and output projections of self-attention in every LLM layer, with rank $r=16$, scaling factor $\alpha_{\mathrm{LoRA}}=32$, and dropout 0.1. 
            The input embeddings and LM output head were also updated. 
            The LoRA updates were merged into the corresponding self-attention projection matrices after this phase. 
            We extended the tokenizer with task-specific special tokens and updated the corresponding embedding rows. 
            For the base LLaMA models, we added the speaker-change token $\langle\mathrm{sc}\rangle$. 
            The instruction-tuned models additionally used
            $\langle\mathrm{pad}\rangle$,
            $\langle\mathrm{bos\_prompt}\rangle$,
            $\langle\mathrm{eos\_prompt}\rangle$,
            $\langle\mathrm{bos\_speech}\rangle$,
            $\langle\mathrm{eos\_speech}\rangle$,
            $\langle\mathrm{bos\_response}\rangle$, and
            $\langle\mathrm{eos\_response}\rangle$
            to delimit the instruction, speech, and response segments.

        \paragraph{Serialized-CTC initialization}
            We initialized the WavLM encoder and LLM decoder from the SOT-backbone checkpoint trained in phase~(a) (Section~\ref{sec:training_details}--a), rather than directly from the public pretrained checkpoints, and attached the separator and $K$ talker-specific CTC heads.
            The existing SOT pathway, including WavLM and the LLM decoder, was frozen, and only the separator and CTC heads were updated. 
            This phase was optimized exclusively using $\mathcal{L}_{\mathrm{SerCTC}}$ in Eq.~(\ref{serialized-ctc-loss}); no SOT loss was applied.

        \paragraph{Stage~1: residual acoustic access}
            Starting from the serialized-CTC checkpoint, we first trained the cross-attention adapters using the direct stacked residual interface. 
            The speech encoder, mixture-prefix pathway, separator, CTC heads, and dense LLM parameters were frozen. 
            Only the decoder-side cross-attention adapters were updated using $\mathcal{L}_{\mathrm{SOT}}$; the serialized-CTC loss was not used.
            For the sequential gated system, the trained stacked-access checkpoint was then used to initialize the gated interface. 
            The cross-attention parameters and layer-wise gate logits were updated using only $\mathcal{L}_{\mathrm{SOT}}$, while the encoder-side components and dense LLM parameters remained frozen. 
            Each gate logit was initialized as $\gamma_{\ell}=-2$, corresponding to $g_{\ell}\approx0.12$.

        \paragraph{Stage~2: joint low-rank refinement}
            Stage~2 was initialized from the corresponding trained Stage~1 checkpoint.
            All pre-existing parameters were frozen, and new LoRA modules were applied to the query, key, value, and output projections of the cross-attention adapters. 
            The joint-refinement variants additionally applied LoRA to the query, key, value, and output projections of LLM self-attention. 
            The selected configuration used a maximum rank of $r=8$, scaling factor $\alpha_{\mathrm{LoRA}}=4$, LoRA dropout of 0.05, and an initial rank logit of 2.0. 
            Only the newly introduced low-rank parameters were updated, using $\mathcal{L}_{\mathrm{SOT}}$ without a serialized-CTC loss.

    \subsection{Optimization Details}
    \label{sec:optimization_details}

        Training used distributed data parallelism on eight GPUs and full-precision floating-point computation. 
        Unless otherwise specified, we used AdamW with $\beta_1=0.9$, $\beta_2=0.999$, $\epsilon=10^{-8}$, and zero weight decay. 
        The learning rate was linearly warmed up to $3\times10^{-5}$ over the first 400 optimization steps.
        It was then linearly decayed according to the configured scheduler horizon. 
        Gradients were clipped to a maximum norm of 1.0, no gradient accumulation was used, and gradient checkpointing and length-based sample grouping were enabled.

        For the 1B and 3B systems, the per-device batch size was 16 on Libri2Mix and 12 on Libri3Mix, corresponding to global batch sizes of 128 and 96, respectively. 
        For the 8B systems, the per-device batch size was 8, corresponding to a global batch size of 64. 
        All experiments ran for 50 epochs. 
        We used a random seed of 42. 
        Validation loss and greedy-decoding WER were computed every 1,600 steps. 
        Checkpoints were saved at the same interval, and at most five checkpoints were retained. 
        The checkpoint with the lowest recorded validation loss was used for evaluation.

\begin{table}[t]
\centering
\footnotesize
\renewcommand{\arraystretch}{0.9}

\caption{\textbf{WER (\%) with 1B LLaMA decoders on the LibriMix Test Sets.}
Boldface indicates the best LLM-decoding result within each decoder
panel; lower is better.
CA denotes cross-attention.}
\vspace{-5pt}
\label{results1b}

\begin{tabularx}{0.9\columnwidth}{
    @{}
    >{\centering\arraybackslash}p{0.075\columnwidth}
    >{\raggedright\arraybackslash}p{0.37\columnwidth}
    *{4}{>{\centering\arraybackslash}X}
    @{}
}
\toprule[1.5pt]
\multirow{2}{*}{\textbf{ID}}
& \multirow{2}{*}{\textbf{System}}
& \multicolumn{2}{c}{\textbf{Noisy}}
& \multicolumn{2}{c}{\textbf{Clean}} \\
\cmidrule(lr){3-4}
\cmidrule(lr){5-6}
& & \textbf{2Mix} & \textbf{3Mix}
  & \textbf{2Mix} & \textbf{3Mix} \\
\midrule[1pt]

\multicolumn{6}{@{}l}{%
\textbf{(a) Decoder-independent references}} \\

1 & SOT-Conformer
& 17.1 & 28.2 & 7.0 & 14.7 \\

2 & EncSep-LSTM \cite{shi_slt2024}
& 15.9 & 26.5 & 7.3 & 13.4 \\

3 & GEncSep-LSTM \cite{shi_slt2024}
& 15.3 & 25.7 & 6.8 & 14.3 \\

\midrule

\multicolumn{6}{@{}l}{%
\textbf{(b) LLaMA-1B Base}} \\

\multicolumn{6}{@{}l}{%
\itshape \textbf{Prefix and encoder references}} \\

4 & LLM-SOT
& 11.0 & 38.9 & 4.6 & 20.4 \\

5 & Stacked mixture CA
& 11.1 & 34.5 & 5.5 & 29.7 \\

\rowcolor{baselinegray}
6 & Serialized CTC
& 10.8 & 21.6 & 5.0 & 11.0 \\

\addlinespace[1pt]
\multicolumn{6}{@{}l}{%
\itshape \textbf{Static CTC-derived prompts}} \\

7 & AcousticPrompt
& 10.5 & 36.7 & 5.3 & 19.2 \\

8 & TokenPrompt
& 34.4 & 83.2 & 19.0 & 76.5 \\

9 & HybridPrompt (SOP)
& 10.2 & 37.4 & \textbf{3.9} & 19.0 \\

\addlinespace[1pt]
\multicolumn{6}{@{}l}{%
\itshape \textbf{Persistent serialized-memory access}} \\

10 & Stacked access
& 10.3 & 23.9 & 5.9 & 11.6 \\

11 & Gated access
& 9.0 & \textbf{20.5} & 4.1 & \textbf{9.2} \\

\rowcolor{proposedgray}
12.D & \quad $+$ Joint LoRA
& \textbf{8.8} & 20.8 & 4.2 & 9.5 \\

\midrule

\multicolumn{6}{@{}l}{%
\textbf{(c) LLaMA-1B Instruct}} \\

\multicolumn{6}{@{}l}{%
\itshape \textbf{Prefix and encoder references}} \\

13 & LLM-SOT
& 9.1 & 32.3 & 3.7 & 19.1 \\

14 & Stacked mixture CA
& 11.7 & 33.9 & 5.5 & 27.4 \\

\rowcolor{baselinegray}
15 & Serialized CTC
& 10.1 & 20.5 & 4.6 & 10.2 \\

\addlinespace[1pt]
\multicolumn{6}{@{}l}{%
\itshape \textbf{Static CTC-derived prompts}} \\

16 & AcousticPrompt
& 9.5 & 31.3 & 3.5 & 17.1 \\

17 & HybridPrompt (SOP)
& 8.4 & 31.3 & \textbf{3.1} & 17.4 \\

\addlinespace[1pt]
\multicolumn{6}{@{}l}{%
\itshape \textbf{Persistent serialized-memory access}} \\

18 & Stacked access
& 11.1 & 21.0 & 5.5 & 15.0 \\

19 & Gated access
& 7.6 & \textbf{17.4} & 3.2 & 14.1 \\

\rowcolor{proposedgray}
20.D & \quad $+$ Joint LoRA
& \textbf{7.5} & 18.0 & 3.9 & \textbf{10.8} \\

\bottomrule[1.5pt]
\end{tabularx}
\vspace{-10pt}
\end{table}

\begin{table}[t]
\centering
\footnotesize
\renewcommand{\arraystretch}{0.9}
\setlength{\tabcolsep}{1.8pt}

\caption{\textbf{WER (\%) with 3B LLaMA decoders on the LibriMix Test Sets.}
Boldface indicates the best LLM-decoding result within each decoder
panel; lower is better.}
\vspace{-5pt}
\label{results3b}

\begin{tabularx}{0.9\columnwidth}{
    @{}
    >{\centering\arraybackslash}p{0.075\columnwidth}
    >{\raggedright\arraybackslash}p{0.37\columnwidth}
    *{4}{>{\centering\arraybackslash}X}
    @{}
}
\toprule[1.5pt]
\multirow{2}{*}{\textbf{ID}}
& \multirow{2}{*}{\textbf{System}}
& \multicolumn{2}{c}{\textbf{Noisy}}
& \multicolumn{2}{c}{\textbf{Clean}} \\
\cmidrule(lr){3-4}
\cmidrule(lr){5-6}
& & \textbf{2Mix} & \textbf{3Mix}
  & \textbf{2Mix} & \textbf{3Mix} \\
\midrule[1pt]

\multicolumn{6}{@{}l}{%
\textbf{(a) LLaMA-3B Base}} \\

\multicolumn{6}{@{}l}{%
\itshape \textbf{Prefix and encoder references}} \\

21 & LLM-SOT
& 10.0 & 31.1 & 4.1 & 22.0 \\

22 & Serialized CTC
& 10.0 & 20.7 & 4.7 & 10.3 \\

\addlinespace[1pt]
\multicolumn{6}{@{}l}{%
\itshape \textbf{Persistent serialized-memory access}} \\

23 & Stacked access
& 11.8 & 24.2 & 4.6 & 11.2 \\

24 & Gated access
& 10.1 & 21.2 & \textbf{3.7} & 9.4 \\

\rowcolor{proposedgray}
25.D & \quad $+$ Joint LoRA
& \textbf{9.0} & \textbf{19.4}
& \textbf{3.7} & \textbf{9.0} \\

\midrule

\multicolumn{6}{@{}l}{%
\textbf{(b) LLaMA-3B Instruct}} \\

\multicolumn{6}{@{}l}{%
\itshape \textbf{Prefix and encoder references}} \\

26 & LLM-SOT
& 8.4 & 27.9 & 3.4 & 15.2 \\

27 & Serialized CTC
& 10.2 & 20.9 & 4.6 & 11.4 \\

\addlinespace[1pt]
\multicolumn{6}{@{}l}{%
\itshape \textbf{Persistent serialized-memory access}} \\

28 & Stacked access
& 9.6 & 21.6 & 4.3 & 13.7 \\

29 & Gated access
& 8.5 & 19.0 & 3.2 & 10.0 \\

\rowcolor{proposedgray}
30.D & \quad $+$ Joint LoRA
& \textbf{7.7} & \textbf{17.1}
& \textbf{3.1} & \textbf{8.7} \\

\bottomrule[1.5pt]
\end{tabularx}
\vspace{-10pt}
\end{table}

\begin{table}[t]
\centering
\footnotesize
\renewcommand{\arraystretch}{0.9}
\setlength{\tabcolsep}{1.8pt}

\caption{\textbf{WER (\%) with 8B LLaMA decoders on the LibriMix Test Sets.}
Boldface indicates the best LLM-decoding result within each decoder
panel; lower is better.}
\vspace{-5pt}
\label{results8b}

\begin{tabularx}{0.9\columnwidth}{
    @{}
    >{\centering\arraybackslash}p{0.075\columnwidth}
    >{\raggedright\arraybackslash}p{0.37\columnwidth}
    *{4}{>{\centering\arraybackslash}X}
    @{}
}
\toprule[1.5pt]
\multirow{2}{*}{\textbf{ID}}
& \multirow{2}{*}{\textbf{System}}
& \multicolumn{2}{c}{\textbf{Noisy}}
& \multicolumn{2}{c}{\textbf{Clean}} \\
\cmidrule(lr){3-4}
\cmidrule(lr){5-6}
& & \textbf{2Mix} & \textbf{3Mix}
  & \textbf{2Mix} & \textbf{3Mix} \\
\midrule[1pt]

\multicolumn{6}{@{}l}{%
\textbf{(a) LLaMA-8B Base}} \\

\multicolumn{6}{@{}l}{%
\itshape \textbf{Prefix and encoder references}} \\

31 & LLM-SOT
& 12.9 & 43.8 & 5.4 & 31.0 \\

32 & Serialized CTC
& 10.4 & 23.6 & 4.8 & 11.0 \\

\addlinespace[1pt]
\multicolumn{6}{@{}l}{%
\itshape \textbf{Persistent serialized-memory access}} \\

33 & Stacked access
& 12.7 & 29.7 & 4.8 & 13.0 \\

34 & Gated access
& 10.7 & 27.0 & \textbf{4.0} & 11.7 \\

\rowcolor{proposedgray}
35.D & \quad $+$ Joint LoRA
& \textbf{10.4} & \textbf{24.3}
& 4.1 & \textbf{11.1} \\

\midrule

\multicolumn{6}{@{}l}{%
\textbf{(b) LLaMA-8B Instruct}} \\

\multicolumn{6}{@{}l}{%
\itshape \textbf{Prefix and encoder references}} \\

36 & LLM-SOT
& 11.4 & 34.9 & 4.5 & 21.6 \\

37 & Serialized CTC
& 10.4 & 21.5 & 4.9 & 11.1 \\

\addlinespace[1pt]
\multicolumn{6}{@{}l}{%
\itshape \textbf{Persistent serialized-memory access}} \\

38 & Stacked access
& 10.9 & 26.9 & 4.2 & 12.8 \\

39 & Gated access
& 9.6 & 23.6 & 3.5 & 11.1 \\

\rowcolor{proposedgray}
40.D & \quad $+$ Joint LoRA
& \textbf{9.0} & \textbf{21.7}
& \textbf{3.4} & \textbf{10.1} \\

\bottomrule[1.5pt]
\end{tabularx}
\vspace{-10pt}
\end{table}

\begin{table*}[t]
\centering
\caption{\textbf{WER (\%) for controlled comparisons using LLaMA-3.2-1B-Instruct on the LibriMix test sets.}
IDs~42--44 shared the serialized-CTC initialization.
IDs~41 and 45 formed the matched static--persistent comparison using the
same downsampled and projected acoustic feature. 
IDs~46, 47, and 19 used the same stacked-memory checkpoint for the prefill-only and dual-path ablations. 
CA denotes cross-attention; stacked CA used direct residual addition, whereas gated CA used the learned sigmoid-scaled residual in
Eq.~(\ref{eq:retrieval_residual_interfaces}).}
\label{tab:controlled_1b}
\vspace{-5pt}

\footnotesize
\renewcommand{\arraystretch}{1.0}
\setlength{\tabcolsep}{3.7pt}

\begin{tabular}{@{}c l l l c c c c@{}}
\toprule[1.5pt]
\multirow{2}{*}{\textbf{ID}}
& \multirow{2}{*}{\textbf{System}}
& \multirow{2}{*}{\textbf{Static prefix}}
& \multirow{2}{*}{\textbf{Decoder-side memory}}
& \multicolumn{2}{c}{\textbf{Noisy}}
& \multicolumn{2}{c}{\textbf{Clean}} \\
\cmidrule(lr){5-6}
\cmidrule(lr){7-8}
& & & &
\textbf{2Mix} & \textbf{3Mix}
& \textbf{2Mix} & \textbf{3Mix} \\
\midrule[1pt]

\multicolumn{8}{@{}l}{%
\textbf{A. Static information control}\quad
\textit{Are acoustic cues sufficient when delivered only through
the initial prefix?}} \\

13 & LLM-SOT
& $[\mathbf{E}_{\mathrm{inst}},\mathbf{H}_p]$
& None
& 9.1 & 32.3 & 3.7 & 19.1 \\

16 & AcousticPrompt
& $[\mathbf{E}_{\mathrm{inst}},\mathbf{E}_{\mathrm{aco}}]$
& None
& 9.5 & 31.3 & 3.5 & 17.1 \\

17 & HybridPrompt (SOP)
& $[\mathbf{E}_{\mathrm{inst}},\mathbf{E}_{\mathrm{hyb}}]$
& None
& 8.4 & 31.3 & 3.1 & 17.4 \\

41 & Matched tensor / static prefix
& $[\mathbf{E}_{\mathrm{inst}},\mathbf{H}_p,
       \mathbf{E}_{\mathrm{aco}}]$
& None
& 8.8 & 31.9 & 3.5 & 18.2 \\

\midrule

\multicolumn{8}{@{}l}{%
\textbf{B. Memory-source and residual-interface controls}\quad
\textit{How do memory source and residual interface affect
persistent access?}} \\

42 & Mixture memory / stacked access
& $[\mathbf{E}_{\mathrm{inst}},\mathbf{H}_p]$
& Stacked CA on $\mathbf{M}_{\mathrm{mix}}$
& 11.7 & 42.1 & 11.1 & 33.0 \\

43 & Mixture memory / gated access
& $[\mathbf{E}_{\mathrm{inst}},\mathbf{H}_p]$
& Gated CA on $\mathbf{M}_{\mathrm{mix}}$
& 8.1 & 25.6 & 3.3 & 14.8 \\

18 & Serialized memory / stacked access
& $[\mathbf{E}_{\mathrm{inst}},\mathbf{H}_p]$
& Stacked CA on $\mathbf{M}_{\mathrm{ser}}$
& 11.1 & 21.0 & 5.5 & 15.0 \\

44 & Serialized memory / gated access (CTC init.)
& $[\mathbf{E}_{\mathrm{inst}},\mathbf{H}_p]$
& Gated CA on $\mathbf{M}_{\mathrm{ser}}$
& 7.9 & 24.8 & 3.4 & 12.3 \\

19 & Serialized memory / gated access (seq. init.)
& $[\mathbf{E}_{\mathrm{inst}},\mathbf{H}_p]$
& Gated CA on $\mathbf{M}_{\mathrm{ser}}$
& 7.6 & 17.4 & 3.2 & 14.1 \\

\midrule

\multicolumn{8}{@{}l}{%
\textbf{C. Static versus persistent access}\quad
\textit{Does decoder-time access help beyond static or
prefill-only access?}} \\

41 & Matched tensor / static prefix
& $[\mathbf{E}_{\mathrm{inst}},\mathbf{H}_p,
       \mathbf{E}_{\mathrm{aco}}]$
& None
& 8.8 & 31.9 & 3.5 & 18.2 \\

45 & Matched tensor / gated access
& $[\mathbf{E}_{\mathrm{inst}},\mathbf{H}_p]$
& Gated CA on $\mathbf{E}_{m}$
& 8.6 & 27.7 & 3.2 & 15.9 \\

46 & Serialized memory / prefill-only access
& $[\mathbf{E}_{\mathrm{inst}},\mathbf{H}_p]$
& Gated CA on $\mathbf{M}_{\mathrm{ser}}$ (prefill only)
& 8.7 & 30.1 & 3.4 & 17.4 \\

19 & Serialized memory / gated access (seq. init.)
& $[\mathbf{E}_{\mathrm{inst}},\mathbf{H}_p]$
& Gated CA on $\mathbf{M}_{\mathrm{ser}}$
& 7.6 & 17.4 & 3.2 & 14.1 \\

\midrule

\multicolumn{8}{@{}l}{%
\textbf{D. Dual-path conditioning}\quad
\textit{Does the mixture prefix complement persistent
talker-structured memory?}} \\

19 & Serialized memory / gated access (seq. init.)
& $[\mathbf{E}_{\mathrm{inst}},\mathbf{H}_p]$
& Gated CA on $\mathbf{M}_{\mathrm{ser}}$
& 7.6 & 17.4 & 3.2 & 14.1 \\

47 & Serialized memory / gated access (no $\mathbf{H}_p$)
& $[\mathbf{E}_{\mathrm{inst}}]$
& Gated CA on $\mathbf{M}_{\mathrm{ser}}$
& 23.4 & 31.0 & 16.3 & 25.4 \\

\bottomrule[1.5pt]
\end{tabular}
\vspace{-10pt}
\end{table*}

\begin{table}[t]
\centering
\footnotesize
\renewcommand{\arraystretch}{0.9}
\setlength{\tabcolsep}{3.0pt}

\caption{\textbf{Effect of the WavLM feature layer.}
Test-set WER (\%) was reported when the acoustic module used representations from different layers of WavLM-Large. 
Layer~24 denotes the current final-layer configuration under the matched Stage~1 setup.
Lower is better.}
\vspace{-5pt}
\label{tab:wavlm_layer_ablation}

\begin{tabular*}{0.9\columnwidth}{
    @{}
    l
    @{\extracolsep{\fill}}
    c c c c
    @{}
}
\toprule[1.5pt]
\multirow{2}{*}{\textbf{WavLM layer}}
& \multicolumn{2}{c}{\textbf{Noisy}}
& \multicolumn{2}{c}{\textbf{Clean}} \\
\cmidrule(lr){2-3}
\cmidrule(lr){4-5}
& \textbf{2Mix} & \textbf{3Mix}
& \textbf{2Mix} & \textbf{3Mix} \\
\midrule[1pt]

Layer 12
& 31.0 & 56.1 & 19.0 & 41.8 \\

Layer 18
& 15.2 & 38.0 & 7.3 & 36.7 \\

Layer 24 (Exp.--44)
& \textbf{7.9} & \textbf{24.8}
& \textbf{3.4} & \textbf{12.3} \\

\bottomrule[1.5pt]
\end{tabular*}
\vspace{-10pt}
\end{table}

\section{Experimental Results and Analysis}
\label{sec:exp_analysis}

    \subsection{Overall Performance and Talker-Count Scaling}
    \label{sec:analysis_overall}

        Table~\ref{results1b} presents the detailed comparison between static prefix conditioning, encoder-only serialized CTC, and persistent decoder-side memory access. 
        The conventional LLM-SOT systems performed well on Libri2Mix but degraded substantially as the talker count increased. 
        For the Instruct model, the WER increased from 9.1\% to 32.3\% under the noisy condition and from 3.7\% to 19.1\% under the clean condition when moving from Libri2Mix to Libri3Mix. 
        The Base model exhibited the same pattern, increasing from 11.0\% to 38.9\% and from 4.6\% to 20.4\%, respectively.
        Instruction tuning improved the LLM-SOT baseline in all four conditions, but did not resolve its sensitivity to the number of talkers.

        The serialized-CTC references showed substantially less degradation from two to three talkers. 
        For example, the 1B-Instruct serialized-CTC system obtained 10.1\% and 20.5\% WER on noisy Libri2Mix and Libri3Mix, respectively, compared with 9.1\% and 32.3\% for LLM-SOT. 
        Thus, although serialized CTC was not consistently better on Libri2Mix, its talker-specific encoder representations remained considerably more robust as the number of speakers increased.

        Although derived from the same serialized-CTC representations, static prompts yielded only modest improvements. 
        For example, HybridPrompt reduced noisy Libri3Mix WER from 38.9\% to 37.4\% for the Base model and from 32.3\% to 31.3\% for the Instruct model.
        Static prompting was more competitive on Libri2Mix, where HybridPrompt obtained the best clean-condition LLM-decoding result in both decoder panels. 
        By contrast, TokenPrompt performed poorly, indicating that a greedily decoded CTC token sequence did not provide an effective acoustic conditioning signal in this configuration.

        Providing repeated decoder-side access to these representations produced the largest improvements on Libri3Mix. 
        For the Base model, gated access reduced WER from 38.9\% to 20.5\% on noisy Libri3Mix and from 20.4\% to 9.2\% on clean Libri3Mix. 
        For the Instruct model, gated access reduced noisy Libri3Mix WER from 32.3\% to 17.4\%, while Stage~2 refinement reduced clean Libri3Mix WER from 19.1\% to 10.8\%. 
        These gains were substantially larger than those obtained by the static CTC-derived prompts, supporting the benefit of retaining direct memory access during autoregressive generation.
        However, Stage~2 refinement did not improve every condition relative to Stage~1.
        The main gain was therefore attributed to gated persistent memory access, with low-rank refinement providing additional improvements only in selected conditions.

        Tables~\ref{results3b} and~\ref{results8b} show that the benefit over the corresponding LLM-SOT baseline extended to larger decoders. 
        For the 3B-Instruct model, Stage~2 reduced noisy and clean Libri3Mix WER from 27.9\% and 15.2\% to 17.1\% and 8.7\%, respectively.
        For the 8B models, the noisy Libri3Mix WER decreased from 43.8\% to 24.3\% with the Base backbone and from 34.9\% to 21.7\% with the Instruct backbone.
        The corresponding clean Libri3Mix results decreased from 31.0\% to 11.1\% and from 21.6\% to 10.1\%.

        Absolute performance was nevertheless not monotonic with decoder size. 
        On noisy Libri3Mix, the LLM-SOT results for the 1B, 3B, and 8B models were 38.9\%, 31.1\%, and 43.8\% for the Base variants and 32.3\%, 27.9\%, and 34.9\% for the Instruct variants.
        Moreover, the 8B Stage~2 systems remained slightly worse than their serialized-CTC references on this condition. 
        The scaling experiments therefore established improvements over the matched same-scale LLM-SOT baselines, rather than showing that increasing the LLM size necessarily improved multi-talker recognition or that LLM decoding uniformly outperformed encoder-only serialized CTC.

    \subsection{Controlled Memory and Interface Comparisons}
    \label{sec:analysis_controlled_grounding}

        Table~\ref{tab:controlled_1b} organizes the controlled experiments according to the hypothesis examined by each comparison. 
        Panel~A showed that CTC-derived talker information could improve recognition when supplied as a static prefix.
        However, the static configurations remained relatively weak on noisy Libri3Mix, suggesting that providing the acoustic cues only during prefill was insufficient for the more difficult three-talker condition.

        Panel~B examined the memory source and residual interface.
        IDs~42--44 shared the same per-corpus serialized-CTC initialization and 50-epoch optimization budget. 
        With mixture memory fixed, replacing direct residual addition (ID~42) with gated integration (ID~43) substantially reduced WER in all four conditions, indicating that the gate stabilized the introduction of decoder-side acoustic information.
        With the gated interface fixed, onset-ordered talker memory (ID~44) outperformed mixture memory (ID~43) in both noisy conditions and on clean Libri3Mix, while their clean Libri2Mix results were comparable. 
        The advantage of talker-structured memory was therefore more evident for three-talker recognition. 
        ID~18 was retained as the previously published direct-access reference, but it was not used as a strictly budget-matched control for IDs~42--44. 
        Similarly, ID~19 was reported separately as the sequentially initialized Stage~1 reference.

        Panel~C directly compared static, prefill-only, and persistent access.
        IDs~41 and 45 used the same downsampled and projected acoustic tensor, differing only in whether it was concatenated to the static prefix or accessed through gated cross-attention throughout decoding.
        Persistent access reduced WER in all four conditions, with larger improvements on the three-talker sets. 
        The comparison between prefill-only access (ID~46) and persistent access (ID~19), which shared the same stacked-memory initialization and optimization budget, showed the same trend. 
        In particular, retaining memory access during autoregressive generation substantially improved noisy three-talker recognition. 
        These results demonstrated the benefit of making the memory available during generation, but did not imply that every generated token necessarily relied on it.

        Finally, Panel~D evaluated the role of the conventional mixture prefix.
        Removing $\mathbf{H}_p$ while retaining the complete persistent-memory pathway (ID~47) substantially degraded all four conditions relative to the full dual-path system (ID~19). 
        Thus, serialized talker memory did not replace the global mixture prefix; the two conditioning pathways provided complementary information.

\begin{table}[t]
\centering
\renewcommand{\arraystretch}{0.9}
\setlength{\tabcolsep}{2.0pt}
\caption{\textbf{WER (\%) for Stage~2 low-rank refinement ablation on Librimix test sets.} 
Each block started from the indicated Stage~1 checkpoint.
All variants applied LoRA to the cross-attention projections; SA indicates additional LoRA on selected self-attention projections. 
N2/C2 denote the noisy and clean conditions for Libri2Mix; N3/C3 denote the noisy and clean conditions for Libri3Mix.}
\label{tab:stage2_sweep}
\vspace{-5pt}

\begin{tabularx}{0.9\linewidth}{c c c c Y Y Y Y}
\toprule[1.5pt]
\multirow{2}{*}{\textbf{Result ID}}
& \multirow{2}{*}{\textbf{SA}}
& \multicolumn{2}{c}{\textbf{LoRA}}
& \multicolumn{4}{c}{\textbf{Test WER}} \\
\cmidrule(lr){3-4}
\cmidrule(lr){5-8}
& & $r$ & $\alpha_{\mathrm{LoRA}}$
& \textbf{N2} & \textbf{N3}
& \textbf{C2} & \textbf{C3} \\
\midrule[1pt]

\multicolumn{8}{l}{
    \textbf{LLaMA-1B Base}
    \hfill \emph{Stage~1 init.: 11}
} \\
12.A & No  & 4  & 4 & 9.3 & 21.7 & 5.3 & 10.0 \\
12.B & No  & 8  & 2 & 9.4 & 22.0 & 5.3 & 10.2 \\
12.C & No  & 8  & 4 & 9.3 & 21.7 & 5.4 & 10.0 \\
\rowcolor{proposedgray}
12.D & Yes & 8  & 4
& \textbf{8.8} & \textbf{20.8}
& \textbf{4.2} & \textbf{9.5} \\
12.E & No  & 8  & 8 & 9.2 & 21.3 & 5.3 & 10.0 \\
12.F & No  & 16 & 8 & 9.2 & 21.2 & 5.2 & 9.8 \\
\midrule

\multicolumn{8}{l}{
    \textbf{LLaMA-1B Instruct}
    \hfill \emph{Stage~1 init.: 19}
} \\
20.A & No  & 4  & 4 & 8.0 & 18.7 & 4.3 & 14.3 \\
20.B & No  & 8  & 2 & 8.2 & 18.8 & 4.2 & 14.5 \\
20.C & No  & 8  & 4 & 8.1 & 18.8 & 4.0 & 14.1 \\
\rowcolor{proposedgray}
20.D & Yes & 8  & 4
& \textbf{7.5} & \textbf{18.0}
& \textbf{3.9} & \textbf{10.8} \\
20.E & No  & 8  & 8 & 8.0 & 18.7 & 4.0 & 13.6 \\
20.F & No  & 16 & 8 & 8.1 & 18.6 & 4.0 & 13.9 \\
\midrule

\multicolumn{8}{l}{
    \textbf{LLaMA-3B Base}
    \hfill \emph{Stage~1 init.: 24}
} \\
25.A & No  & 4  & 4 & 9.1 & 20.1 & 3.8 & 9.3 \\
25.B & No  & 8  & 2 & 9.3 & 20.3 & 3.8 & 9.4 \\
25.C & No  & 8  & 4 & 9.2 & 20.3 & 3.8 & 9.3 \\
\rowcolor{proposedgray}
25.D & Yes & 8  & 4
& \textbf{9.0} & \textbf{19.4}
& \textbf{3.7} & \textbf{9.0} \\
25.E & No  & 8  & 8 & 9.0 & 20.1 & 3.9 & 9.2 \\
25.F & No  & 16 & 8 & 9.0 & 19.8 & 3.8 & 9.1 \\
\midrule

\multicolumn{8}{l}{
    \textbf{LLaMA-3B Instruct}
    \hfill \emph{Stage~1 init.: 29}
} \\
30.A & No  & 4  & 4 & 8.0 & 17.8 & 3.4 & 9.2 \\
30.B & No  & 8  & 2 & 8.0 & 18.0 & 3.3 & 9.1 \\
30.C & No  & 8  & 4 & 8.0 & 17.9 & 3.4 & 9.2 \\
\rowcolor{proposedgray}
30.D & Yes & 8  & 4
& \textbf{7.7} & \textbf{17.1}
& \textbf{3.1} & \textbf{8.7} \\
30.E & No  & 8  & 8 & 8.0 & 18.0 & 3.5 & 10.2 \\
30.F & No  & 16 & 8 & 8.0 & 17.9 & 3.5 & 10.1 \\
\midrule

\multicolumn{8}{l}{
    \textbf{LLaMA-8B Base}
    \hfill \emph{Stage~1 init.: 34}
} \\
35.A & No  & 4  & 4 & 10.4 & 24.7 & 3.6 & 11.5 \\
35.B & No  & 8  & 2 & 10.5 & 24.7 & 4.2 & 11.5 \\
35.C & No  & 8  & 4 & 10.4 & 24.7 & 4.2 & 11.4 \\
\rowcolor{proposedgray}
35.D & Yes & 8  & 4
& \textbf{10.4} & \textbf{24.3}
& \textbf{4.1} & \textbf{11.1} \\
35.E & No  & 8  & 8 & 10.4 & 24.7 & 4.2 & 11.5 \\
35.F & No  & 16 & 8 & 10.5 & 24.7 & 4.2 & 11.4 \\
\midrule

\multicolumn{8}{l}{
    \textbf{LLaMA-8B Instruct}
    \hfill \emph{Stage~1 init.: 39}
} \\
40.A & No  & 4  & 4 & 9.4 & 21.6 & 4.2 & 10.4 \\
40.B & No  & 8  & 2 & 9.2 & 21.6 & 3.7 & 10.4 \\
40.C & No  & 8  & 4 & 9.4 & 21.7 & 3.6 & 10.4 \\
\rowcolor{proposedgray}
40.D & Yes & 8  & 4
& \textbf{9.0} & \textbf{21.7}
& \textbf{3.4} & \textbf{10.1} \\
40.E & No  & 8  & 8 & 9.6 & 21.8 & 3.6 & 10.6 \\
40.F & No  & 16 & 8 & 9.9 & 21.9 & 3.6 & 10.5 \\
\bottomrule[1.5pt]
\end{tabularx}
\vspace{-10pt}
\end{table}

\begin{table}[t]
\renewcommand{\arraystretch}{0.9}
\caption{Comparison between the proposed method and existing methods
on the LibriMix datasets (270 hours for Libri2Mix and 186 hours for Libri3Mix, without additional data augmentation).
``Orig.'' indicates whether the reported results used the original LibriMix evaluation mixtures without additional onset offsets. 
``Without LLM decoding at inference'' refers only to the inference
architecture.}
\label{table:method_comparison}

\footnotesize
\vspace{-5pt}
\centering
\setlength{\tabcolsep}{3.0pt}

\begin{tabular}{ll|c|cc|cc}
\toprule[1.5pt]
\multicolumn{2}{c|}{\multirow{2}{*}{\textbf{REF}}}
& \multirow{2}{*}{\shortstack{\textbf{Orig.}\\\textbf{no off.}}}
& \multicolumn{2}{c|}{\textbf{Libri2Mix}}
& \multicolumn{2}{c}{\textbf{Libri3Mix}} \\
\cline{4-7}
& & & \textbf{Dev} & \textbf{Test}
& \textbf{Dev} & \textbf{Test} \\

\midrule[1pt]

\multicolumn{1}{c|}{\multirow{20}{*}{\textbf{Noisy}}}
& \multicolumn{6}{c}{
\textbf{Without an LLM decoder}} \\
\cline{2-7}

\multicolumn{1}{c|}{}
& Conformer-based SOT \cite{shi_slt2024}
& \ding{55} & 19.4 & 17.1 & 30.5 & 28.2 \\

\multicolumn{1}{c|}{}
& TSE-V-Whisper \cite{10389752}
& \ding{51} & -- & 12.0 & -- & -- \\

\multicolumn{1}{c|}{}
& GEncSep \cite{shi_slt2024}
& \ding{55} & 17.2 & 15.0 & 28.0 & 25.9 \\

\multicolumn{1}{c|}{}
& \quad CTC-only inference
& \ding{55} & 16.8 & 14.6 & 25.7 & 23.6 \\

\multicolumn{1}{c|}{}
& UME \cite{shakeel2025unifying}
& \ding{51} & -- & 19.6 & -- & 27.1 \\

\multicolumn{1}{c|}{}
& HCM \cite{kashiwagi2025hypothesis}
& \ding{51} & -- & 18.4 & -- & 36.3 \\

\multicolumn{1}{c|}{}
& SoloSpeech \cite{wang2025solospeech}
& \ding{51} & -- & 18.0 & -- & -- \\


\cline{2-7}

\multicolumn{1}{c|}{}
& \multicolumn{6}{c}{\textbf{With an LLM decoder}} \\
\cline{2-7}

\multicolumn{1}{c|}{}
& SOT-LLaMA-1B (ID~4)
& \ding{55} & 12.6 & 11.0 & 39.8 & 38.9 \\
\multicolumn{1}{c|}{}
& SOT-LLaMA-1B (ID~4)
& \ding{51} & 18.2 & 17.6 & 60.2 & 61.1 \\

\multicolumn{1}{c|}{}
& SOP-LLaMA-1B (ID~9)
& \ding{55} & 11.8 & 10.2 & 38.7 & 37.4 \\
\multicolumn{1}{c|}{}
& SOP-LLaMA-1B (ID~9)
& \ding{51} & 17.2 & 15.9 & 58.8 & 59.7 \\

\multicolumn{1}{c|}{}
& SOT-LLaMA-3B (ID~21)
& \ding{55} & 11.2 & 10.0 & 32.9 & 31.1 \\
\multicolumn{1}{c|}{}
& SOT-LLaMA-3B (ID~21)
& \ding{51} & 15.7 & 15.2 & 55.4 & 56.6 \\

\multicolumn{1}{c|}{}
& SOP-LLaMA-3B \cite{shi2025serialized}
& \ding{55} & 10.5 & 9.2 & 29.3 & 28.1 \\

\cline{2-7}

\multicolumn{1}{c|}{}
& ID-20.D
& \ding{55} & 8.4 & 7.5 & 19.2 & 18.0 \\
\multicolumn{1}{c|}{}
& ID-20.D
& \ding{51} & 13.3 & 13.1 & 43.3 & 45.3 \\

\multicolumn{1}{c|}{}
& ID-30.D
& \ding{55} & 8.6 & 7.7 & 18.8 & 17.1 \\
\multicolumn{1}{c|}{}
& ID-30.D
& \ding{51} & 14.0 & 13.5 & 41.0 & 42.6 \\

\midrule

\multicolumn{1}{c|}{\multirow{24}{*}{\textbf{Clean}}}
& \multicolumn{6}{c}{
\textbf{Without an LLM decoder}} \\
\cline{2-7}

\multicolumn{1}{c|}{}
& Conformer-based SOT \cite{shi_slt2024}
& \ding{55} & 6.8 & 7.0 & 15.0 & 14.7 \\

\multicolumn{1}{c|}{}
& W2V-Sidecar-ft. \cite{10095295}
& \ding{51} & 7.7 & 8.1 & -- & -- \\

\multicolumn{1}{c|}{}
& WavLM-CLN \cite{10097139}
& \ding{51} & 7.1 & 7.6 & -- & -- \\

\multicolumn{1}{c|}{}
& C-HuBERT Large \cite{10096630}
& \ding{51} & 6.6 & 7.8 & -- & -- \\

\multicolumn{1}{c|}{}
& GEncSep \cite{shi_slt2024}
& \ding{55} & 6.4 & 6.6 & 13.3 & 13.1 \\

\multicolumn{1}{c|}{}
& \quad CTC-only inference
& \ding{55} & 6.0 & 6.1 & 12.7 & 12.5 \\

\multicolumn{1}{c|}{}
& UME \cite{shakeel2025unifying}
& \ding{51} & -- & 6.4 & -- & 15.9 \\

\multicolumn{1}{c|}{}
& HCM \cite{kashiwagi2025hypothesis}
& \ding{51} & -- & 8.2 & -- & 21.5 \\

\multicolumn{1}{c|}{}
& MUSE-TSASR \cite{seong2025enhancing}
& \ding{51} & -- & 11.0 & -- & 10.5 \\

\multicolumn{1}{c|}{}
& SoloSpeech \cite{wang2025solospeech}
& \ding{51} & -- & 15.0 & -- & -- \\

\multicolumn{1}{c|}{}
& Whisper-Sidecar \cite{meng24c_interspeech}
& \ding{51} & -- & 4.7 & -- & 16.8 \\

\cline{2-7}

\multicolumn{1}{c|}{}
& \multicolumn{6}{c}{\textbf{With an LLM decoder}} \\
\cline{2-7}

\multicolumn{1}{c|}{}
& SOT-LLaMA-1B (ID~4)
& \ding{55} & 4.5 & 4.6 & 20.7 & 20.4 \\
\multicolumn{1}{c|}{}
& SOT-LLaMA-1B (ID~4)
& \ding{51} & 8.4 & 8.6 & 46.1 & 48.4 \\

\multicolumn{1}{c|}{}
& SOP-LLaMA-1B (ID~9)
& \ding{55} & 3.8 & 3.9 & 19.9 & 19.0 \\
\multicolumn{1}{c|}{}
& SOP-LLaMA-1B (ID~9)
& \ding{51} & 7.7 & 8.1 & 45.0 & 46.6 \\

\multicolumn{1}{c|}{}
& SOT-LLaMA-3B (ID~21)
& \ding{55} & 4.2 & 4.1 & 21.8 & 22.0 \\
\multicolumn{1}{c|}{}
& SOT-LLaMA-3B (ID~21)
& \ding{51} & 7.4 & 7.7 & 47.7 & 50.4 \\

\multicolumn{1}{c|}{}
& SOP-LLaMA-3B
& \ding{55} & 3.5 & 3.6 & 17.0 & 16.5 \\

\cline{2-7}

\multicolumn{1}{c|}{}
& ID-20.D
& \ding{55} & 3.8 & 3.9 & 9.8 & 10.8 \\
\multicolumn{1}{c|}{}
& ID-20.D
& \ding{51} & 7.8 & 8.2 & 28.6 & 31.6 \\

\multicolumn{1}{c|}{}
& ID-30.D
& \ding{55} & 3.0 & 3.1 & 8.6 & 8.7 \\
\multicolumn{1}{c|}{}
& ID-30.D
& \ding{51} & 6.5 & 6.9 & 28.1 & 30.4 \\

\bottomrule[1.5pt]
\end{tabular}

\vspace{-10pt}
\end{table}

    \subsection{Intermediate-Layer WavLM Front Ends}
    \label{sec:intermediate_wavlm}

        The original system constructed its talker-specific representations from the final output of the 24-layer WavLM-Large encoder.
        This design required all WavLM layers to remain active before the separator constructed the serialized acoustic memory. 
        Motivated by intermediate-layer sidecar architectures \cite{10095295}, we investigated whether the upper WavLM layers were necessary for this purpose. 
        Intermediate representations might preserve sufficient local acoustic and talker-discriminative information before the stronger contextualization introduced by the final layers.
        If such representations were sufficient, the remaining upper encoder blocks could be removed, reducing the active encoder parameters and acoustic-front-end computation.

        We therefore constructed two WavLM-Large variants truncated after Transformer layer 12 or 18. 
        For a truncation depth $\ell\in\{12,18\}$, the output of layer $\ell$ was supplied directly to the existing separator and subsequent acoustic-conditioning modules. 
        Layers $\ell+1$ through 24 were removed from both training and inference. 
        This differed from merely tapping an intermediate representation while retaining the full encoder: truncation produced an actual reduction in deployment-time encoder depth and computation. 
        The separator architecture was held fixed across the two variants so that the experiment primarily examined the effect of the WavLM feature depth and the removal of the upper encoder blocks. 
        Both variants used LLaMA-3.2-1B-Instruct as the language decoder.

        The separator and serialized-CTC heads were initialized from the same serialized-CTC checkpoint used to initialize Exp.~44 and were further fine-tuned in these experiments. 
        The retained WavLM layers, temporal adapter, \texttt{enc\_to\_dec\_proj}, separator, and serialized-CTC heads were trainable. 
        For the LLM, LoRA was applied to the query, key, value, and output projections of self-attention. 
        The input embeddings, language-model output head and the remaining dense LLM parameters were frozen. 
        The models were optimized using the training objective in Eq.~(\ref{eq:decoder_adaptation_loss}).

        The results in Table~\ref{tab:wavlm_layer_ablation} show that the performance improved monotonically as more WavLM layers were retained. 
        Retaining all 24 WavLM layers produced the best result in every condition, with larger gains in the three-talker conditions. 
        One possible explanation was that deeper WavLM representations provided contextual or linguistic information useful for recognizing overlapping speech. 
        In this sense, the term ``acoustic memory'' described the speech-derived source of the representations and did not imply that they contain only low-level acoustic information. 
        However, all retained WavLM layers were trainable, so increasing encoder depth also increased the number of trainable parameters and the capacity for task-specific adaptation. 
        The present comparison therefore could not isolate the contribution of richer contextual representations from that of increased trainable capacity. 
        We interpreted these results as an empirical accuracy--cost trade-off for encoder truncation under the evaluated training protocol.

    \subsection{Stage~2 Joint Low-Rank Refinement}
    \label{sec:analysis_stage2_refinement}

        Stage~2 was a downstream refinement of the trained gated cross-attention system rather than evidence for the architectural effect of gating.
        The complete sweep in Table~\ref{tab:stage2_sweep} separated the effect of adding self-attention LoRA from changes in cross-attention LoRA rank and scaling.
        For the matched $r=8$, $\alpha_{\mathrm{LoRA}}=4$ pairs, joint refinement lowered or matched test WER in every displayed cell. 
        Thus, the evidence supported joint low-rank refinement as a useful second-stage optimization, but not a uniform claim that its benefit grows with model size. 
        Across that sweep, changing cross-attention-only LoRA capacity produced relatively small and non-monotonic differences; therefore, larger rank alone did not explain the gains of the joint configuration.

\begin{table}[t]
\centering
\footnotesize
\renewcommand{\arraystretch}{0.9}
\setlength{\tabcolsep}{2.8pt}

\caption{\textbf{WER and error decomposition (\%) on the derived LibriCSS evaluation subsets.}
$N_{\mathrm{ref}}$ is the number of reference words.
Sub, Del, and Ins denote substitution, deletion, and insertion
rates normalized by $N_{\mathrm{ref}}$.
0S and 0L denote non-overlapping speech with short and long
inter-utterance silences, respectively.
}
\vspace{-5pt}

\label{tab:libricss_wer}

\resizebox{0.48\textwidth}{!}{%
\begin{tabular}{@{}l r | rrrr | rrrr | r@{}}
\toprule[1.5pt]

\multirow{2}{*}{\textbf{Condition}}
& \multirow{2}{*}{$\mathbf{N_{\mathrm{ref}}}$}
& \multicolumn{4}{c|}{\textbf{LLM-SOT}}
& \multicolumn{4}{c|}{\textbf{Proposed}}
& \multirow{2}{*}{\textbf{$\Delta$WER}} \\

\cmidrule(lr){3-6}
\cmidrule(lr){7-10}

& & \textbf{Sub} & \textbf{Del} & \textbf{Ins} & \textbf{WER}
& \textbf{Sub} & \textbf{Del} & \textbf{Ins} & \textbf{WER}
& \\

\midrule[1pt]

\multicolumn{11}{@{}l}{\textbf{(a) Two-talker clips}} \\

0S
& 9,907
& 3.07 & 1.78 & 6.89 & 11.74
& 3.49 & 0.77 & 2.45 & \textbf{6.71}
& 5.03 \\

0L
& 6,069
& 3.48 & 1.65 & 7.28 & 12.41
& 3.59 & 0.86 & 2.17 & \textbf{6.62}
& 5.79 \\

OV10
& 10,436
& 4.77 & 1.95 & 9.70 & 16.42
& 5.06 & 1.69 & 5.59 & \textbf{12.33}
& 4.09 \\

OV20
& 10,576
& 5.57 & 2.45 & 12.16 & 20.18
& 6.45 & 2.82 & 10.68 & \textbf{19.95}
& 0.23 \\

OV30
& 13,073
& 7.02 & 2.56 & 19.61 & 29.19
& 7.99 & 3.13 & 15.09 & \textbf{26.21}
& 2.98 \\

OV40
& 12,518
& 9.95 & 3.36 & 22.92 & \textbf{36.23}
& 10.59 & 4.49 & 22.34 & 37.42
& -1.19 \\

\addlinespace[1pt]

\textbf{Overall}
& \textbf{62,579}
& 6.02 & 2.39 & 14.15 & 22.56
& 6.62 & 2.51 & 10.96 & \textbf{20.09}
& 2.47 \\

\midrule[1pt]

\multicolumn{11}{@{}l}{\textbf{(b) Three-talker clips}} \\

0S
& 3,580
& 7.43 & 3.55 & 24.55 & 35.53
& 8.04 & 7.01 & 7.37 & \textbf{22.43}
& 13.10 \\

\textbf{0L}
& \textbf{813}
& 8.12 & 0.74 & 41.21 & 50.06
& 6.03 & 4.18 & 5.29 & \textbf{15.50}
& \textbf{34.56} \\

OV10
& 3,794
& 8.30 & 4.19 & 22.85 & 35.35
& 6.88 & 8.30 & 12.23 & \textbf{27.41}
& 7.94 \\

OV20
& 3,863
& 10.77 & 6.76 & 20.48 & 38.00
& 10.17 & 7.30 & 14.94 & \textbf{32.41}
& 5.59 \\

OV30
& 5,917
& 11.22 & 7.35 & 24.88 & 43.45
& 11.53 & 6.66 & 16.83 & \textbf{35.02}
& 8.43 \\

OV40
& 5,752
& 15.79 & 7.67 & 27.73 & 51.18
& 15.33 & 9.70 & 24.04 & \textbf{49.08}
& 2.10 \\

\addlinespace[1pt]

\textbf{Overall}
& \textbf{23,719}
& 11.11 & 6.02 & 25.04 & 42.17
& 10.77 & 7.73 & 15.71 & \textbf{34.22}
& 7.95 \\

\bottomrule[1.5pt]
\end{tabular}%
}
\vspace{-10pt}
\end{table}

\begin{table*}[t]
\centering
\scriptsize
\renewcommand{\arraystretch}{0.9}
\setlength{\tabcolsep}{2.2pt}

\caption{\textbf{Inference efficiency and deployment parameter
footprint.}
Each slash-separated entry reports Libri2Mix/Libri3Mix.
All systems were evaluated on an NVIDIA A100-SXM4-80GB GPU using
BF16 precision, batch size one, greedy decoding, and no CPU offloading.
Each result was measured over 200 utterances after 10 warm-up
utterances and averaged over three repetitions.
All LLM-decoding systems used self-attention KV caching; the
memory-access systems additionally cached the projected keys and values
of the fixed acoustic memory at every cross-attention layer.
E2E latency includes host-to-device transfer, WavLM encoding, memory
construction, memory-KV projection, prompt prefill, and autoregressive
decoding, but excludes model loading and disk I/O.
Audio throughput is measured in input-audio seconds per wall-clock
second.
Active inference parameters exclude the serialized-CTC heads for
LLM-decoding systems, whereas these heads remain active for
CTC-only inference.}
\vspace{-10pt}
\label{tab:inference_efficiency}

\textbf{(a) End-to-end inference efficiency}\par\vspace{2pt}
\begin{tabular*}{0.86\textwidth}{
@{\extracolsep{\fill}}
c l c c c c c c c c c
@{}
}
\toprule[1.5pt]
\textbf{Scale}
& \textbf{System}
& \shortstack{\textbf{Active}\\\textbf{params (B)}}
& \shortstack{\textbf{Added}\\\textbf{params (M)}}
& \shortstack{\textbf{E2E mean}\\\textbf{(ms)}}
& \shortstack{\textbf{E2E P90}\\\textbf{(ms)}}
& \textbf{RTF}
& \shortstack{\textbf{Audio}\\\textbf{throughput}}
& \shortstack{\textbf{Token}\\\textbf{throughput}}
& \shortstack{\textbf{Decode}\\\textbf{ms/token}} \\
\midrule[1pt]

\multirow{5}{*}{\textbf{1B}}
& LLM-SOT
& 1.472
& --
& 747/1094
& 1060/1472
& .083/.100
& 12.04/10.01
& 64.1/63.5
& 15.6/15.7  \\

& Serialized CTC
& .510/.643
& --
& 167/219
& 250/339
& .019/.020
& 53.80/49.96
& --
& --  \\

& Stacked access
& 1.536/1.537
& 64.4/65.8
& 1212/1651
& 1684/2213
& .135/.151
& 7.42/6.63
& 47.7/47.2
& 21.0/21.2 \\

& Gated access
& 1.536/1.537
& 64.4/65.8
& 1167/1680
& 1664/2239
& .130/.153
& 7.71/6.52
& 46.0/45.9
& 21.7/21.8  \\

& Gated $+$ Joint LoRA
& 1.537/1.538
& 65.4/66.9
& 1312/1925
& 1894/2504
& .146/.176
& 6.86/5.69
& 40.5/39.9
& 24.7/25.1  \\

\midrule

\multirow{5}{*}{\textbf{3B}}
& LLM-SOT
& 3.449
& --
& 1239/1844
& 1781/2496
& .138/.168
& 7.26/5.94
& 38.0/37.6
& 26.3/26.6 \\

& Serialized CTC
& .510/.643
& --
& 181/209
& 282/293
& .020/.019
& 49.73/52.52
& --
& --  \\

& Stacked access
& 3.581/3.583
& 131.8/133.2
& 1795/2759
& 2581/3695
& .200/.252
& 5.01/3.97
& 28.3/27.1
& 35.4/36.9  \\

& Gated access
& 3.581/3.583
& 131.8/133.2
& 1936/2818
& 2764/3741
& .215/.257
& 4.65/3.89
& 26.2/26.2
& 38.1/38.2 \\

& Gated $+$ Joint LoRA
& 3.584/3.585
& 134.1/135.5
& 2313/3440
& 3285/4650
& .257/.314
& 3.89/3.18
& 21.7/21.1
& 46.1/47.3 \\

\midrule

\multirow{5}{*}{\textbf{8B}}
& LLM-SOT
& 8.268
& --
& 1407/2123
& 1995/2875
& .156/.194
& 6.40/5.16
& 33.4/32.5
& 30.0/30.7  \\

& Serialized CTC
& .510/.643
& --
& 166/197
& 230/267
& .018/.018
& 54.21/55.64
& --
& --  \\

& Stacked access
& 8.450/8.452
& 182.3/183.8
& 2280/3435
& 3324/4697
& .253/.314
& 3.95/3.19
& 21.8/21.6
& 45.8/46.2  \\

& Gated access
& 8.450/8.452
& 182.3/183.8
& 2093/3069
& 3028/4046
& .233/.280
& 4.30/3.57
& 23.8/24.1
& 42.0/41.5 \\

& Gated $+$ Joint LoRA
& 8.454/8.455
& 185.5/186.9
& 2435/3486
& 3474/4663
& .271/.318
& 3.69/3.14
& 20.4/20.9
& 49.1/47.9  \\

\bottomrule[1.5pt]
\end{tabular*}

\vspace{5pt}
\textbf{(b) Module-level parameter breakdown (not phase-specific trainable counts)}\par\vspace{2pt}
\begin{tabular*}{0.86\textwidth}{
@{\extracolsep{\fill}}
c c c c c c c
@{}
}
\toprule[1.2pt]
\textbf{Scale}
& \shortstack{\textbf{WavLM}\\\textbf{(M)}}
& \shortstack{\textbf{Adapter +}\\\textbf{projector (M)}}
& \shortstack{\textbf{Separator}\\\textbf{(M)}}
& \shortstack{\textbf{CA projections}\\\textbf{+ gate (M)}}
& \shortstack{\textbf{Stage-2 LoRA}\\\textbf{weights (M)}}
& \shortstack{\textbf{CTC heads}\\\textbf{(M)}} \\
\midrule

1B
& 315.5
& 21.0
& 13.9/15.3
& 50.5
& 1.0
& 262.9/394.4 \\

3B
& 315.5
& 22.0
& 13.9/15.3
& 117.9
& 2.3
& 262.9/394.4 \\

8B
& 315.5
& 23.1
& 13.9/15.3
& 168.5
& 3.1
& 262.9/394.4 \\

\bottomrule[1.2pt]

\end{tabular*}

\vspace{1pt}
\begin{minipage}{0.94\textwidth}
\scriptsize
Panel~(b) reports module-level parameter counts, not the number
of parameters trained simultaneously.
Trainable and frozen components vary across training phases,
as detailed in Section~V-D. 
Slash-separated separator and CTC-head counts correspond to the two-/three-talker systems. 
The CTC heads are used only during training for the LLM-decoding systems and are removed for deployment; they remain active only in the
Serialized CTC system. 
The Stage~2 LoRA counts denote separately stored low-rank parameters before merging.
\end{minipage}
\vspace{-10pt}
\end{table*}

    \subsection{Comparison with Existing Systems}
    \label{sec:analysis_comparison}

        Table~\ref{table:method_comparison} compares the proposed systems with existing methods under both noisy and clean acoustic conditions.
        The ``Orig.'' column explicitly distinguishes results obtained on the original LibriMix evaluation mixtures from those obtained using the additional onset-offset configurations. 
        The original Libri2Mix and Libri3Mix test sets had approximate overlap ratios of 49.9\% and 64.7\%, respectively; 
        three-speaker overlap accounted for 36.0\% of speech-active frames in Libri3Mix. 
        Overlap ratios follow the definition in Section~V-A.
        On the offset evaluation sets, the proposed systems performed particularly well on Libri3Mix. 
        To examine whether this improvement transferred to the standard evaluation mixtures, we additionally re-evaluated selected 1B and 3B models on the original LibriMix test sets without additional onset offsets. 
        The degradation was especially large on Libri3Mix. 
        This gap indicated that the current systems were sensitive to the onset configuration and that performance on the offset mixtures did not directly predict performance on the original LibriMix mixtures. 
        In particular, the absence of additional offsets produced more simultaneous or weakly separated onsets, making the onset-based serialization order more ambiguous.

To further examine ordering ambiguity on the original LibriMix mixtures, we analyzed the decoding outputs of five systems under both clean and noisy conditions.
All source waveforms started at sample zero, while the reference transcripts followed metadata source-index order.
After removing prompt wrappers, we split each hypothesis at speaker-change tokens and, for hypotheses with exactly $K$ nonempty segments, found the one-to-one assignment to reference transcripts that minimized summed word-level edit distance.
Among hypotheses with a unique minimum-cost assignment, agreement with source-index order ranged from 49.3\% to 50.5\% on Libri2Mix and from 16.0\% to 17.8\% on Libri3Mix.
The observed permutations were approximately evenly distributed, indicating no strong preference for the metadata source-index order.
These statistics covered 67.1--99.9\% of mixtures; hypotheses with missing or extra segments or tied assignments were excluded. 
These results were consistent with increased ordering ambiguity on the original LibriMix mixtures.
However, waveform start did not necessarily coincide with actual speech onset, and reliable onset annotations were unavailable.
The analysis therefore did not establish whether the generated order followed true speech onset or other acoustic cues, and transcript-based assignments could also have been affected by recognition errors.

    \subsection{Cross-Domain Evaluation on LibriCSS}
    \label{sec:analysis_libricss}

        Table~\ref{tab:libricss_wer} evaluates transfer to far-field room recordings without retraining on LibriCSS~\cite{9053426}.
        Across both the two- and three-talker subsets, the overall WER improvement was driven primarily by fewer insertion errors. 
        For the two-talker subset, the overall substitution, deletion, and insertion rates changed from 6.02\%, 2.39\%, and 14.15\% for LLM-SOT to 6.62\%, 2.51\%, and 10.96\% for the proposed system, respectively. 
        Thus, the WER reduction from 22.56\% to 20.09\% resulted from a 3.19-point reduction in insertions, despite small increases in substitutions and deletions. 
        The same pattern was more pronounced for the three-talker subset. 
        The overall substitution rate decreased slightly from 11.11\% to 10.77\%, while the insertion rate decreased substantially from 25.04\% to 15.71\%.
        The deletion rate, however, increased from 6.02\% to 7.73\%. 
        Together, these changes reduced the overall WER from 42.17\% to 34.22\%.

        The three-talker 0L condition provided the clearest example of this insertion-dominated behavior. 
        For LLM-SOT, the substitution, deletion, and insertion rates were 8.12\%, 0.74\%, and 41.21\%, respectively, showing that most of its 50.06\% WER originated from insertions. 
        The proposed system reduced the insertion rate to 5.29\% and the substitution rate to 6.03\%, although its deletion rate increased to 4.18\%, resulting in a substantially lower WER of 15.50\%. 
        This error pattern was consistent with reduced over-generation and more stable serialization in the presence of long pauses, rather than an improvement attributable solely to acoustic modeling. 
        Because the three-talker 0L subset contained only 24 clips from seven sessions, this observation should nevertheless be interpreted cautiously.

        The advantage became limited under the most severe overlap condition. 
        For two-talker OV40, insertions decreased only slightly from 22.92\% to 22.34\%, while substitutions and deletions increased from 9.95\% and 3.36\% to 10.59\% and 4.49\%, respectively. 
        Consequently, the proposed system was slightly worse than LLM-SOT (37.42\% versus 36.23\% WER). 
        For three-talker OV40, the insertion rate decreased from 27.73\% to 24.04\%, but the deletion rate increased from 7.67\% to 9.70\%; the resulting WER improvement was therefore modest (49.08\% versus 51.18\%). 
        We interpreted these cross-domain results as evidence that serialized acoustic-memory access primarily mitigates insertion-dominated failures, particularly under long pauses, while robustness to severe overlap remains limited.

    \subsection{Inference Efficiency and Deployment Cost}
    \label{sec:analysis_efficiency}

        Table~\ref{tab:inference_efficiency} reports the inference efficiency and active deployment parameters of the 1B, 3B, and 8B systems. 
        The LLM-SOT baselines achieved RTFs ranging from 0.083 to 0.194, whereas the gated memory-access systems ranged from 0.130 to 0.280. 
        The corresponding audio throughputs remained between 3.57 and 7.71 seconds of input audio per second of processing. 
        Relative to the same-scale LLM-SOT baseline, the end-to-end latency of the stacked and gated systems increased by approximately $1.45\times$--$1.62\times$. 
        For the 1B models, gated access increased weighted decoding latency from 15.6/15.7 to 21.7/21.8\,ms per token for Libri2Mix/Libri3Mix. 
        The corresponding changes were from 26.3/26.6 to 38.1/38.2\,ms per token for 3B and from 30.0/30.7 to 42.0/41.5\,ms per token for 8B. 
        Thus, most of the remaining overhead arose during repeated cross-attention-based decoding rather than from the one-time construction of the memory cache.

        The higher end-to-end latency on Libri3Mix was also partly explained by the input and output distributions. 
        The evaluated Libri3Mix utterances were longer on average and contained approximately 1,641 memory positions and 67 generated tokens, compared with 898 memory positions and 45 generated tokens for Libri2Mix.  
        In contrast, the weighted per-token latency changed only modestly between the two corpora for a fixed system. 
        The Libri2Mix--Libri3Mix difference in end-to-end latency therefore reflected both the longer serialized memory and the larger number of generated tokens.
        The additional memory required by the cross-attention K/V cache was moderate: approximately 28/51\,MB for 1B, 49/90\,MB for 3B, and 56/103\,MB for 8B on the two-/three-talker subsets.

        In parameter terms, gated access added 64--66M active parameters to the 1B baseline, 132--133M to the 3B baseline, and 182--184M to the 8B baseline. 
        These additions corresponded to approximately 4.4\%, 3.8\%, and 2.2\% of the respective LLM-SOT deployment models. 
        Consequently, the proposed interface was parameter-efficient relative to the size of the LLM backbone, but it was not computationally lightweight.
        Repeated attention over a long serialized acoustic memory introduced
        non-negligible latency and memory overhead.

    \subsection{Streaming Limitation}
    \label{sec:streaming_limitation}

        The proposed method is currently limited to offline recognition.
        Serialized acoustic memory is constructed from the complete input
        segment and remains fixed during autoregressive decoding.
        Accordingly, ``persistent access'' refers only to the decoder's repeated
        access to this complete segment-level memory.
        It does not imply that memory extracted from an initial chunk remains
        sufficient when subsequent audio arrives.

        For strict streaming recognition, a fixed memory constructed from the
        initial chunk cannot contain acoustic evidence from future chunks and is
        therefore insufficient.
        The memory would need to be updated as new chunks arrive.
        Such an extension would also require a causal or chunkwise encoder,
        consistent tracking of talker branches across chunks, and a
        streaming-compatible SOT emission policy.
        These components are not included or evaluated in the present system,
        and we therefore do not claim streaming capability.

        A more direct extension is buffered long-form transcription, in which
        a complete memory is constructed and held fixed within each fully
        observed window and then reconstructed for the next window.
        Adjacent windows can include an overlapping region, which can be used
        to maintain cross-window speaker consistency through PIT-style
        permutation matching between their talker-specific representations or
        predicted token sequences.
        After determining the speaker correspondence, the hypotheses can be
        stitched by removing duplicated tokens in the overlap.
        A simple strategy is to discard tokens from the later window whose
        timestamps fall within the overlapping region.
        Although the autoregressive decoder does not directly produce token
        timestamps, approximate timing information can be obtained from the
        corresponding CTC alignment.
        We leave the implementation and evaluation of this buffered long-form
        extension for future work.

\begin{figure*}[t]
    \centering
    \includegraphics[width=\textwidth]{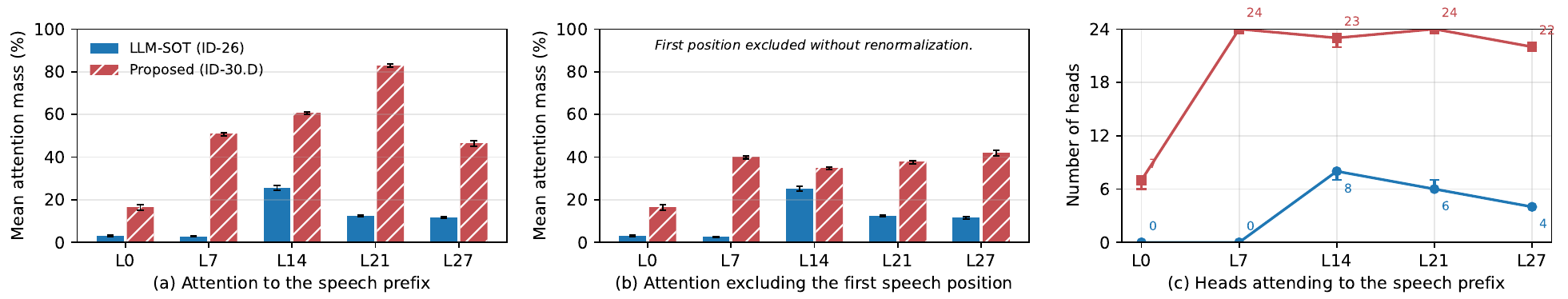}
    \vspace{-20pt}
    \caption{
    Teacher-forced decoder self-attention to the speech prefix
    in LLM-SOT (ID~26) and the proposed model (ID~30.D),
    evaluated on offset noisy Libri3Mix test sets at five selected layers. 
    Statistics were computed over lexical target tokens;
    error bars denote sample-level bootstrap 95\% confidence intervals.
    }
    \label{fig:speech_prefix_attention}
\vspace{-10pt}
\end{figure*}

    \subsection{Speech-Prefix Attention Across Decoder Depth}
    \label{sec:speech_prefix_attention}

        To examine how the two systems access the conventional acoustic
        prefix, we analyzed decoder self-attention in LLM-SOT (ID--26)
        and the proposed model (ID--30.D) in Figure~\ref{fig:speech_prefix_attention}. 
        The speech prefix $\mathbf{H}_p$ denotes the projected acoustic
        representations embedded in the decoder input sequence, distinct
        from the external serialized memory accessed through cross-attention.
        We inspected five pre-specified layers,
        L0, L7, L14, L21, and L27 (zero-based indexing), covering
        different depths of the 28-layer decoder (3B-Instruct Model). 
        For each of the 24 query heads in a selected layer, we summed
        the self-attention probabilities assigned to valid speech-prefix
        positions at each lexical-token prediction step.
        We averaged these values over lexical tokens within each sample
        and then equally across samples, excluding control tokens. 
        The proposed model assigned more attention to the speech prefix at all five inspected layers.
        This difference was not fully explained by attention to the first speech position: after excluding that position, the proposed model still exhibited higher attention mass throughout the inspected depths. 
        The first-position contribution was nevertheless substantial at L14 and L21, highlighting the importance of distinguishing
        total prefix attention from attention to its remaining positions. 
        The difference also extended across heads. 
        At L7 and L21, all 24 heads in the proposed model exceeded the total-mass threshold, compared with zero and six heads in LLM-SOT, respectively.
        These observations indicated that, in the inspected layers, the proposed system distributed attention to the conventional speech prefix across more heads, despite retaining an additional external-memory pathway.

\section{Conclusion}

    In this work, we investigated why conventional LLM-based serialized output training performs well for two-talker mixtures but degrades substantially as the number of competing talkers increases.
    Continuous CTC-derived prompts were more effective than discrete CTC hypotheses, but their limited improvements on Libri3Mix showed that enriching the initial prefix alone is insufficient. 
    We therefore proposed SOT-aligned serialized acoustic memory with persistent decoder-side access. 
    Talker-specific representations are organized in onset order and retained as external memory, while gated residual cross-attention makes them directly available during autoregressive generation. 
    Here, ``persistent'' refers to the continued availability of the memory, rather than guaranteed reliance on it at every decoding step. 
    The controlled comparisons clarify the contributions of this design. 
    Gated integration consistently improved over direct stacked cross-attention in the reported configurations. 
    The matched static--persistent and prefill-only comparisons further showed that retaining access during generation is more effective than presenting the same acoustic information only in the initial context. 
    Removing the conventional mixture prefix caused substantial degradation, demonstrating that the global mixture representation and talker-structured memory provide complementary information. 
    Across the 1B, 3B, and 8B backbones, persistent memory access substantially improved the corresponding same-scale LLM-SOT baselines, with the largest gains generally observed on three-talker mixtures. 
    Joint low-rank refinement provided additional improvements in selected conditions, with its most consistent gains observed for the 3B models, but did not uniformly improve every Stage~1 result. 
    Performance was also not monotonic with decoder size. 
    In particular, the 8B systems did not surpass encoder-only serialized CTC on noisy Libri3Mix despite their greater inference cost. 
    The results therefore support persistent acoustic access rather than a general claim that larger LLM decoders necessarily improve multi-talker ASR. 
    The optimized implementation caches both the LLM self-attention states and the projected keys and values of the fixed acoustic memory. 
    Although this substantially reduces the cost of the original uncached implementation, repeated cross-attention still introduces non-negligible latency. 
    The proposed interface is therefore parameter-efficient relative to the LLM backbone, but should not be characterized as computationally lightweight. 
    Serialized CTC remains the preferable accuracy--efficiency trade-off in some conditions. 
    Evaluation on the derived fixed-talker LibriCSS subsets showed transfer to far-field recordings without retraining, with the overall gains arising primarily from fewer insertion errors. 
    However, improvements became limited under severe overlap. 
    Improving robustness to ambiguous onsets, unknown talker counts, severe overlap, and streaming or long-form input remains important future work.

\bibliographystyle{IEEEtran}
\bibliography{mybib}

@article{9814838,
  author  = {Chen, Sanyuan and Wang, Chengyi and Chen, Zhengyang and Wu, Yu and Liu, Shujie and Chen, Zhuo and Li, Jinyu and Kanda, Naoyuki and Yoshioka, Takuya and Xiao, Xiong and Wu, Jian and Zhou, Long and Ren, Shuo and Qian, Yanmin and Qian, Yao and Wu, Jian and Zeng, Michael and Yu, Xiangzhan and Wei, Furu},
  title   = {{WavLM}: Large-Scale Self-Supervised Pre-Training for Full Stack Speech Processing},
  journal = {IEEE J. Sel. Topics Signal Process.},
  year    = {2022},
  volume  = {16},
  number  = {6},
  pages   = {1505--1518},
  doi     = {10.1109/JSTSP.2022.3188113},
}

@inproceedings{10096630,
  author    = {Fazel-Zarandi, Maryam and Hsu, Wei-Ning},
  title     = {Cocktail {HuBERT}: Generalized Self-Supervised Pre-Training for Mixture and Single-Source Speech},
  booktitle = {Proc. ICASSP},
  year      = {2023},
  pages     = {1--5},
  doi       = {10.1109/ICASSP49357.2023.10096630},
}

@inproceedings{shi2025serialized,
  author    = {Hao Shi and Yusuke Fujita and Tomoya Mizumoto and Lianbo Liu and Atsushi Kojima and Yui Sudo},
  title     = {Serialized Output Prompting for Large Language Model-Based Multi-Talker Speech Recognition},
  booktitle = {Proc. ASRU},
  year      = {2025},
  pages     = {1--8},
  doi       = {10.1109/ASRU65441.2025.11434801},
}

@article{10542371,
  author  = {Shi, Hao and Mimura, Masato and Kawahara, Tatsuya},
  title   = {Waveform-Domain Speech Enhancement Using Spectrogram Encoding for Robust Speech Recognition},
  journal = {IEEE/ACM Trans. Audio, Speech, Language Process.},
  year    = {2024},
  volume  = {32},
  pages   = {3049--3060},
  doi     = {10.1109/TASLP.2024.3407511},
}

@inproceedings{shi2024advancing,
  author    = {Shi, Mohan and Jin, Zengrui and Xu, Yaoxun and Xu, Yong and Zhang, Shi-Xiong and Wei, Kun and Shao, Yiwen and Zhang, Chunlei and Yu, Dong},
  title     = {Advancing Multi-Talker {ASR} Performance With Large Language Models},
  booktitle = {Proc. SLT},
  year      = {2024},
  pages     = {14--21},
}

@inproceedings{shi_slt2024,
  author    = {Shi, Hao and Gao, Yuan and Ni, Zhaoheng and Kawahara, Tatsuya},
  title     = {Serialized Speech Information Guidance with Overlapped Encoding Separation for Multi-Speaker Automatic Speech Recognition},
  booktitle = {Proc. SLT},
  year      = {2024},
  pages     = {198--204},
}

@inproceedings{meng2024large,
  author    = {Meng, Lingwei and Hu, Shujie and Kang, Jiawen and Li, Zhaoqing and Wang, Yuejiao and Wu, Wenxuan and Wu, Xixin and Liu, Xunying and Meng, Helen},
  title     = {Large Language Model Can Transcribe Speech in Multi-Talker Scenarios with Versatile Instructions},
  booktitle = {Proc. ICASSP},
  year      = {2025},
  pages     = {1--5},
}

@article{6732927,
  author  = {Li, Jinyu and Deng, Li and Gong, Yifan and Haeb-Umbach, Reinhold},
  title   = {An Overview of Noise-Robust Automatic Speech Recognition},
  journal = {IEEE/ACM Trans. Audio, Speech, Language Process.},
  year    = {2014},
  volume  = {22},
  number  = {4},
  pages   = {745--777},
  doi     = {10.1109/TASLP.2014.2304637},
}

@article{6423821,
  author  = {Deng, Li and Li, Xiao},
  title   = {Machine Learning Paradigms for Speech Recognition: An Overview},
  journal = {IEEE Trans. Audio, Speech, Language Process.},
  year    = {2013},
  volume  = {21},
  number  = {5},
  pages   = {1060--1089},
  doi     = {10.1109/TASL.2013.2244083},
}

@inproceedings{8682822,
  author    = {Chang, Xuankai and Qian, Yanmin and Yu, Kai and Watanabe, Shinji},
  title     = {End-to-end Monaural Multi-speaker {ASR} System without Pretraining},
  booktitle = {Proc. ICASSP},
  year      = {2019},
  pages     = {6256--6260},
  doi       = {10.1109/ICASSP.2019.8682822},
}

@inproceedings{8461893,
  author    = {Settle, Shane and Roux, Jonathan Le and Hori, Takaaki and Watanabe, Shinji and Hershey, John R.},
  title     = {End-to-End Multi-Speaker Speech Recognition},
  booktitle = {Proc. ICASSP},
  year      = {2018},
  pages     = {4819--4823},
  doi       = {10.1109/ICASSP.2018.8461893},
}

@article{6739096,
  author  = {Narayanan, Arun and Wang, DeLiang},
  title   = {Investigation of Speech Separation as a Front-End for Noise Robust Speech Recognition},
  journal = {IEEE/ACM Trans. Audio, Speech, Language Process.},
  year    = {2014},
  volume  = {22},
  number  = {4},
  pages   = {826--835},
  doi     = {10.1109/TASLP.2014.2305833},
}

@article{8369155,
  author  = {Wang, DeLiang and Chen, Jitong},
  title   = {Supervised Speech Separation Based on Deep Learning: An Overview},
  journal = {IEEE/ACM Trans. Audio, Speech, Language Process.},
  year    = {2018},
  volume  = {26},
  number  = {10},
  pages   = {1702--1726},
  doi     = {10.1109/TASLP.2018.2842159},
}

@article{7979557,
  author  = {Kolbæk, Morten and Yu, Dong and Tan, Zheng-Hua and Jensen, Jesper},
  title   = {Multitalker Speech Separation With Utterance-Level Permutation Invariant Training of Deep Recurrent Neural Networks},
  journal = {IEEE/ACM Trans. Audio, Speech, Language Process.},
  year    = {2017},
  volume  = {25},
  number  = {10},
  pages   = {1901--1913},
  doi     = {10.1109/TASLP.2017.2726762},
}

@inproceedings{9054328,
  author    = {Tripathi, Anshuman and Lu, Han and Sak, Hasim},
  title     = {End-To-End Multi-Talker Overlapping Speech Recognition},
  booktitle = {Proc. ICASSP},
  year      = {2020},
  pages     = {6129--6133},
  doi       = {10.1109/ICASSP40776.2020.9054328},
}

@inproceedings{kanda20b_interspeech,
  author    = {Kanda, Naoyuki and Gaur, Yashesh and Wang, Xiaofei and Meng, Zhong and Yoshioka, Takuya},
  title     = {{Serialized Output Training for End-to-End Overlapped Speech Recognition}},
  booktitle = {Proc. Interspeech},
  year      = {2020},
  pages     = {2797--2801},
}

@inproceedings{hu2021lora,
  author    = {Hu, Edward J. and Shen, Yelong and Wallis, Phillip and Allen-Zhu, Zeyuan and Li, Yuanzhi and Wang, Shean and Wang, Lu and Chen, Weizhu},
  title     = {{LoRA}: Low-Rank Adaptation of Large Language Models},
  booktitle = {Proc. ICLR},
  year      = {2022},
}

@article{cosentino2020librimix,
  author  = {Cosentino, Joris and Pariente, Manuel and Cornell, Samuele and Deleforge, Antoine and Vincent, Emmanuel},
  title   = {{LibriMix}: An open-source dataset for generalizable speech separation},
  journal = {arXiv preprint arXiv:2005.11262},
  year    = {2020},
}

@inproceedings{7178964,
  author    = {Panayotov, Vassil and Chen, Guoguo and Povey, Daniel and Khudanpur, Sanjeev},
  title     = {{LibriSpeech}: An {ASR} corpus based on public domain audio books},
  booktitle = {Proc. ICASSP},
  year      = {2015},
  pages     = {5206--5210},
  doi       = {10.1109/ICASSP.2015.7178964},
}

@inproceedings{radford2023robust,
  author    = {Radford, Alec and Kim, Jong Wook and Xu, Tao and Brockman, Greg and McLeavey, Christine and Sutskever, Ilya},
  title     = {Robust speech recognition via large-scale weak supervision},
  booktitle = {Proc. ICML},
  year      = {2023},
  pages     = {28492--28518},
}

@inproceedings{10095295,
  author    = {Meng, L. and Kang, J. and Cui, M. and Wang, Y. and Wu, X. and Meng, H.},
  title     = {A Sidecar Separator Can Convert A Single-Talker Speech Recognition System to A Multi-Talker One},
  booktitle = {Proc. ICASSP},
  year      = {2023},
  pages     = {1--5},
}

@inproceedings{10097139,
  author    = {Huang, Z. and Raj, D. and García, P. and Khudanpur, S.},
  title     = {Adapting Self-Supervised Models to Multi-Talker Speech Recognition Using Speaker Embeddings},
  booktitle = {Proc. ICASSP},
  year      = {2023},
  pages     = {1--5},
}

@inproceedings{10389752,
  author    = {Zhang, W. and Yang, L. and Qian, Y.},
  title     = {Exploring Time-Frequency Domain Target Speaker Extraction For Causal and Non-Causal Processing},
  booktitle = {Proc. ASRU},
  year      = {2023},
  pages     = {1--6},
}

@inproceedings{cornell24_chime,
  author    = {Samuele Cornell and Tae Jin Park and He Huang and Christoph Boeddeker and Xuankai Chang and Matthew Maciejewski and Matthew S Wiesner and Paola Garcia and Shinji Watanabe},
  title     = {The {CHiME-8} {DASR} Challenge for Generalizable and Array Agnostic Distant Automatic Speech Recognition and Diarization},
  booktitle = {Proc. CHiME},
  year      = {2024},
  pages     = {1--6},
  doi       = {10.21437/CHiME.2024-1},
}

@article{Cornell2025RecentTrends,
  author  = {S. Cornell and C. Boeddeker and T. Park and H. Huang and D. Raj and M. Wiesner and Y. Masuyama and X. Chang and Z.-Q. Wang and S. Squartini and P. Garcia and S. Watanabe},
  title   = {Recent Trends in Distant Conversational Speech Recognition: A Review of {CHiME-7} and {CHiME-8} {DASR} Challenges},
  journal = {Comput. Speech Lang.},
  year    = {2026},
  volume  = {97},
  pages   = {101901},
  doi     = {10.1016/j.csl.2025.101901},
}

@inproceedings{LiangYLGZ0023,
  author    = {Yuhao Liang and Fan Yu and Yangze Li and Pengcheng Guo and Shiliang Zhang and Qian Chen and Lei Xie},
  title     = {{BA-SOT}: Boundary-Aware Serialized Output Training for Multi-Talker {ASR}},
  booktitle = {Proc. Interspeech},
  year      = {2023},
  pages     = {3487--3491},
}

@inproceedings{shakeel2025unifying,
  author    = {Shakeel, Muhammad and Sudo, Yui and Peng, Yifan and Lin, Chyi-Jiunn and Watanabe, Shinji},
  title     = {Unifying Diarization, Separation, and {ASR} with Multi-Speaker Encoder},
  booktitle = {Proc. ASRU},
  year      = {2025},
  pages     = {1--7},
  doi       = {10.1109/ASRU65441.2025.11434654},
}

@inproceedings{kashiwagi2025hypothesis,
  author    = {Kashiwagi, Yosuke and Futami, Hayato and Tsunoo, Emiru and Arora, Siddhant and Watanabe, Shinji},
  title     = {Hypothesis Clustering and Merging: Novel {MultiTalker} Speech Recognition with Speaker Tokens},
  booktitle = {Proc. ICASSP},
  year      = {2025},
  pages     = {1--5},
}

@inproceedings{seong2025enhancing,
  author    = {Seong, Ju-Seok and Choi, Jeong-Hwan and Jeoung, Ye-Rin and Kim, Ilseok and Chang, Joon-Hyuk},
  title     = {Enhancing Target-speaker Automatic Speech Recognition Using Multiple Speaker Embedding Extractors with Virtual Speaker Embedding},
  booktitle = {Proc. Interspeech},
  year      = {2025},
  pages     = {4918--4922},
}

@article{wang2025solospeech,
  author  = {Wang, Helin and Hai, Jiarui and Yang, Dongchao and Chen, Chen and Li, Kai and Peng, Junyi and Thebaud, Thomas and Velazquez, Laureano Moro and Villalba, Jesus and Dehak, Najim},
  title   = {{SoloSpeech}: Enhancing Intelligibility and Quality in Target Speech Extraction Through a Cascaded Generative Pipeline},
  journal = {IEEE Trans. Audio, Speech, Language Process.},
  year    = {2026},
  volume  = {34},
  pages   = {2408--2420},
  doi     = {10.1109/TASLPRO.2026.3682073},
}

@article{fan2025joint,
  author  = {Cunhang Fan and Jiahao Li and Enrui Liu and Jiangyan Yi and Xinhui Li and Ruibo Fu and Zhao Lv},
  title   = {A Joint Training Framework for Noise-Robust Speech Recognition through Multi-Level Feature Fusion},
  journal = {IEEE Trans. Audio, Speech, Language Process.},
  year    = {2025},
  volume  = {33},
  pages   = {4808--4820},
}

@article{fan2025crossmodal,
  author  = {Cunhang Fan and Wang Xiang and Jianhua Tao and Jiangyan Yi and Zhao Lv},
  title   = {Cross-Modal Knowledge Distillation with Multi-Stage Adaptive Feature Fusion for Speech Separation},
  journal = {IEEE Trans. Audio, Speech, Language Process.},
  year    = {2025},
  volume  = {33},
  pages   = {935--948},
}

@inproceedings{li2023prompting,
  author    = {Li, Yuang and Wu, Yu and Li, Jinyu and Liu, Shujie},
  title     = {Prompting Large Language Models for Zero-Shot Domain Adaptation in Speech Recognition},
  booktitle = {Proc. ASRU},
  year      = {2023},
  pages     = {1--8},
  doi       = {10.1109/ASRU57964.2023.10389732},
}

@inproceedings{li24b_interspeech,
  author    = {Li, Yuang and Yu, Jiawei and Zhang, Min and Ren, Mengxin and Zhao, Yanqing and Zhao, Xiaofeng and Tao, Shimin and Su, Jinsong and Yang, Hao},
  title     = {Using Large Language Model for End-to-End Chinese {ASR} and {NER}},
  booktitle = {Proc. Interspeech},
  year      = {2024},
  pages     = {822--826},
  doi       = {10.21437/Interspeech.2024-103},
}

@inproceedings{hirano25_interspeech,
  author    = {Hirano, Yuta and Sakti, Sakriani},
  title     = {{SC-SOT}: Conditioning the Decoder on Diarized Speaker Information for End-to-End Overlapped Speech Recognition},
  booktitle = {Proc. Interspeech},
  year      = {2025},
  pages     = {4923--4927},
  doi       = {10.21437/Interspeech.2025-1243},
}

@inproceedings{kang2025disentangling,
  author    = {Kang, Jiawen and Meng, Lingwei and Cui, Mingyu and Wang, Yuejiao and Wu, Xixin and Liu, Xunying and Meng, Helen},
  title     = {Disentangling Speakers in Multi-Talker Speech Recognition with Speaker-Aware {CTC}},
  booktitle = {Proc. ICASSP},
  year      = {2025},
  pages     = {1--5},
  doi       = {10.1109/ICASSP49660.2025.10888841},
}

@inproceedings{sakuma2025speakerdistinguishable,
  author    = {Sakuma, Asahi and Sato, Hiroaki and Sugano, Ryuga and Kumano, Tadashi and Kawai, Yoshihiko and Ogawa, Tetsuji},
  title     = {Speaker-Distinguishable {CTC}: Learning Speaker Distinction Using {CTC} for Multi-Talker Speech Recognition},
  booktitle = {Proc. Interspeech},
  year      = {2025},
  pages     = {5503--5507},
  doi       = {10.21437/Interspeech.2025-2572},
}

@inproceedings{10446059,
  author    = {Fan, Zhiyun and Dong, Linhao and Zhang, Jun and Lu, Lu and Ma, Zejun},
  title     = {{SA-SOT}: Speaker-Aware Serialized Output Training for Multi-Talker {ASR}},
  booktitle = {Proc. ICASSP},
  year      = {2024},
  pages     = {9986--9990},
  doi       = {10.1109/ICASSP48485.2024.10446059},
}

@inproceedings{kang2024crossspeaker,
  author    = {Kang, Jiawen and Meng, Lingwei and Cui, Mingyu and Guo, Haohan and Wu, Xixin and Liu, Xunying and Meng, Helen},
  title     = {Cross-Speaker Encoding Network for Multi-Talker Speech Recognition},
  booktitle = {Proc. ICASSP},
  year      = {2024},
  pages     = {11986--11990},
  doi       = {10.1109/ICASSP48485.2024.10446249},
}

@inproceedings{yu2024connecting,
  author    = {Yu, Wenyi and Tang, Changli and Sun, Guangzhi and Chen, Xianzhao and Tan, Tian and Li, Wei and Lu, Lu and Ma, Zejun and Zhang, Chao},
  title     = {Connecting Speech Encoder and Large Language Model for {ASR}},
  booktitle = {Proc. ICASSP},
  year      = {2024},
  pages     = {12637--12641},
  doi       = {10.1109/ICASSP48485.2024.10445874},
}

@inproceedings{meng24c_interspeech,
  author    = {Meng, Lingwei and Kang, Jiawen and Wang, Yuejiao and Jin, Zengrui and Wu, Xixin and Liu, Xunying and Meng, Helen},
  title     = {Empowering {Whisper} as a Joint Multi-Talker and Target-Talker Speech Recognition System},
  booktitle = {Proc. Interspeech},
  year      = {2024},
  pages     = {4653--4657},
  doi       = {10.21437/Interspeech.2024-971},
}

@inproceedings{he2025cmt,
  author    = {He, Jiajun and Sawada, Naoki and Miyazaki, Koichi and Toda, Tomoki},
  title     = {{CMT-LLM}: Contextual Multi-Talker {ASR} Utilizing Large Language Models},
  booktitle = {Proc. Interspeech},
  year      = {2025},
  pages     = {2575--2579},
  doi       = {10.21437/Interspeech.2025-943},
}

@article{zhang2020improving,
  author  = {Zhang, Wangyou and Chang, Xuankai and Qian, Yanmin and Watanabe, Shinji},
  title   = {Improving End-to-End Single-Channel Multi-Talker Speech Recognition},
  journal = {IEEE/ACM Trans. Audio, Speech, Language Process.},
  year    = {2020},
  volume  = {28},
  pages   = {1385--1394},
  doi     = {10.1109/TASLP.2020.2988423},
}

@article{raj2023surt2,
  author  = {Raj, Desh and Povey, Daniel and Khudanpur, Sanjeev},
  title   = {{SURT} 2.0: Advances in Transducer-Based Multi-Talker Speech Recognition},
  journal = {IEEE/ACM Trans. Audio, Speech, Language Process.},
  year    = {2023},
  volume  = {31},
  pages   = {3800--3813},
  doi     = {10.1109/TASLP.2023.3318398},
}

@article{qian2018singlechannel,
  author  = {Qian, Yanmin and Chang, Xuankai and Yu, Dong},
  title   = {Single-Channel Multi-Talker Speech Recognition with Permutation Invariant Training},
  journal = {Speech Commun.},
  year    = {2018},
  volume  = {104},
  pages   = {1--11},
  doi     = {10.1016/j.specom.2018.09.003},
  month   = nov,
}

@inproceedings{raj2021integration,
  author    = {Raj, Desh and Denisov, Pavel and Chen, Zhuo and Erdogan, Hakan and Huang, Zili and He, Maokui and Watanabe, Shinji and Du, Jun and Yoshioka, Takuya and Luo, Yi and Kanda, Naoyuki and Li, Jinyu and Wisdom, Scott and Hershey, John R.},
  title     = {Integration of Speech Separation, Diarization, and Recognition for Multi-Speaker Meetings: System Description, Comparison, and Analysis},
  booktitle = {Proc. SLT},
  year      = {2021},
  pages     = {897--904},
  doi       = {10.1109/SLT48900.2021.9383556},
}

@inproceedings{kanda2021speakerattributed,
  author    = {Kanda, Naoyuki and Chang, Xuankai and Gaur, Yashesh and Wang, Xiaofei and Meng, Zhong and Chen, Zhuo and Yoshioka, Takuya},
  title     = {Investigation of End-to-End Speaker-Attributed {ASR} for Continuous Multi-Talker Recordings},
  booktitle = {Proc. SLT},
  year      = {2021},
  pages     = {809--816},
  doi       = {10.1109/SLT48900.2021.9383600},
}

@inproceedings{kanda22_interspeech,
  author    = {Kanda, Naoyuki and Wu, Jian and Wu, Yu and Xiao, Xiong and Meng, Zhong and Wang, Xiaofei and Gaur, Yashesh and Chen, Zhuo and Li, Jinyu and Yoshioka, Takuya},
  title     = {Streaming Multi-Talker {ASR} with Token-Level Serialized Output Training},
  booktitle = {Proc. Interspeech},
  year      = {2022},
  pages     = {3774--3778},
  doi       = {10.21437/Interspeech.2022-7},
}

@inproceedings{kanda21_interspeech,
  author    = {Kanda, Naoyuki and Ye, Guoli and Wu, Yu and Gaur, Yashesh and Wang, Xiaofei and Meng, Zhong and Chen, Zhuo and Yoshioka, Takuya},
  title     = {Large-Scale Pre-Training of End-to-End Multi-Talker {ASR} for Meeting Transcription with Single Distant Microphone},
  booktitle = {Proc. Interspeech},
  year      = {2021},
  pages     = {3430--3434},
  doi       = {10.21437/Interspeech.2021-102},
}

@inproceedings{polok2026grounding,
  author        = {Alexander Polok and Samuele Cornell and Sathvik Udupa and Jan {\v{C}}ernock{\'y} and Shinji Watanabe and Luk{\'a}{\v{s}} Burget},
  title         = {Grounding Spoken {LLMs} in Multi-Speaker Audio via Diarization Conditioning},
  booktitle     = {Proc. Interspeech},
  year          = {2026},
  eprint        = {2606.18134},
  archivePrefix = {arXiv},
  primaryClass  = {eess.AS},
}

@inproceedings{alayrac2022flamingo,
  author    = {Alayrac, Jean-Baptiste and Donahue, Jeff and Luc, Pauline and Miech, Antoine and Barr, Iain and Hasson, Yana and Lenc, Karel and Mensch, Arthur and Millican, Katherine and Reynolds, Malcolm and Ring, Roman and Rutherford, Eliza and Cabi, Serkan and Han, Tengda and Gong, Zhitao and Samangooei, Sina and Monteiro, Marianne and Menick, Jacob L. and Borgeaud, Sebastian and Brock, Andy and Nematzadeh, Aida and Sharifzadeh, Sahand and Bi{\'n}kowski, Miko{\l}aj and Barreira, Ricardo and Vinyals, Oriol and Zisserman, Andrew and Simonyan, Kar{\'e}n},
  title     = {{Flamingo}: A Visual Language Model for Few-Shot Learning},
  booktitle = {Adv. Neural Inf. Process. Syst.},
  year      = {2022},
  volume    = {35},
  pages     = {23716--23736},
  doi       = {10.52202/068431-1723},
}

@inproceedings{zhang2024llamaadapter,
  author    = {Zhang, Renrui and Han, Jiaming and Liu, Chris and Zhou, Aojun and Lu, Pan and Qiao, Yu and Li, Hongsheng and Gao, Peng},
  title     = {{LLaMA-Adapter}: Efficient Fine-Tuning of Large Language Models with Zero-Initialized Attention},
  booktitle = {Proc. ICLR},
  year      = {2024},
}

@inproceedings{li2023blip2,
  author    = {Li, Junnan and Li, Dongxu and Savarese, Silvio and Hoi, Steven},
  title     = {{BLIP}-2: Bootstrapping Language-Image Pre-Training with Frozen Image Encoders and Large Language Models},
  booktitle = {Proc. ICML},
  year      = {2023},
  volume    = {202},
  pages     = {19730--19742},
}

@inproceedings{laskowski12_interspeech,
  author    = {Kornel Laskowski and Mattias Heldner and Jens Edlund},
  title     = {On the Dynamics of Overlap in Multi-Party Conversation},
  booktitle = {Proc. Interspeech},
  year      = {2012},
  pages     = {847--850},
  doi       = {10.21437/Interspeech.2012-191},
}

@inproceedings{he2025scaling,
  author    = {Xiluo He and Alexander Polok and Jes{\'u}s Villalba and Thomas Thebaud and Matthew Maciejewski},
  title     = {Scaling Multi-Talker {ASR} with Speaker-Agnostic Activity Streams},
  booktitle = {Proc. ICASSP},
  year      = {2026},
  doi       = {10.1109/ICASSP55912.2026.11461880},
}

@inproceedings{9053426,
  author    = {Zhuo Chen and Takuya Yoshioka and Liang Lu and Tianyan Zhou and Zhong Meng and Yan Luo and Jian Wu and Xunying Xiao and Jinyu Li},
  title     = {Continuous Speech Separation: Dataset and Analysis},
  booktitle = {Proc. ICASSP},
  year      = {2020},
  pages     = {7284--7288},
  doi       = {10.1109/ICASSP40776.2020.9053426},
}

\end{document}